\documentclass[10pt,onecolumn]{elsarticle}
\usepackage{amssymb,amsmath,theorem} 
\usepackage{times} 
\newtheorem{theorem}{Theorem}

\newtheorem{lemma}{Lemma}

\journal{Annals of Physics}
\begin{document}

\begin{frontmatter}



\title{Deterministic LOCC transformation of three-qubit pure states and entanglement transfer}


\author{Hiroyasu Tajima}

\address{Department of Physics, The University of Tokyo\\ 
4-6-1 Komaba, Meguro, Tokyo, 153-8505, Japan\\
TEL: +81-3-5452-6156 \quad FAX: +81-3-5452-6155\\
E-mail: h-tajima@iis.u-tokyo.ac.jp}

\begin{abstract}
A necessary and sufficient condition of the possibility of a deterministic local operations and classical communication (LOCC) transformation of three-qubit pure states is given.
The condition shows that the three-qubit pure states are a partially ordered set parametrized by  five well-known entanglement parameters and a novel parameter; the five are the concurrences $C_{AB}$, $C_{AC}$, $C_{BC}$, the tangle $\tau_{ABC}$ and the fifth parameter $J_{5}$ of Ref. \cite{18}, while the other new one is the entanglement charge $Q_{\mbox{e}}$.  
The order of the partially ordered set is defined by the possibility of a deterministic LOCC transformation from a state to another state.
In this sense, the present condition is an extension of Nielsen's work \cite{20} to three-qubit pure states.
We also clarify the rules of transfer and dissipation of the entanglement which is caused by  deterministic LOCC transformations.
Moreover, the minimum number of times of measurements to reproduce an arbitrary deterministic LOCC transformation between three-qubit pure states is given.
\end{abstract}


\end{frontmatter}

% \linenumbers

\section{Introduction}
Entanglement is known to be a promising resource which enables us to execute various quantum tasks such as quantum computing, teleportation, superdense coding, $etc.$ \cite{1,2,3,4}. Quantification of the entanglement is a very important subject.

The quantification has been successful for bipartite pure states. 
Many indices such as the concurrence \cite{5,6,7} and the negativity \cite{8} have been proposed. Bennett $et$ $al.$ \cite{9} have proven that all of them can be expressed by the set of the coefficients of the Schmidt decomposition \cite{14,15,16,17}.
Based on the properties of this set, the following have been given: 
\begin{description}
\item[(i)]an explicit necessary and sufficient condition to determine whether an arbitrary bipartite pure state is an entangled state or a separable state;
\item[(i$\hspace{-.1em}$i)]an explicit necessary and sufficient condition to determine whether we can transform an arbitrary bipartite pure state into another arbitrary bipartite pure state with local unitary transformation (LU-equivalence);
\item[(i$\hspace{-.1em}$i$\hspace{-.1em}$i)]an explicit necessary and sufficient condition to determine whether a deterministic LOCC transformation from an arbitrary bipartite pure state to another arbitrary bipartite pure state is executable or not \cite{20};
\item[(i$\hspace{-.1em}$v)]an explicit necessary and sufficient condition to determine whether a stochastic LOCC transformation from an arbitrary bipartite pure state to an arbitrary set of bipartite pure states with arbitrary probability is executable or not \cite{21};
\item[(v)]the fact that copies of an arbitrary partially entangled pure state can be distilled to the Bell states by an LOCC transformation, where the ratio between the copies and the Bell states is proportional to the entanglement entropy of the partially entangled state \cite{9};
\item[(v$\hspace{-.1em}$i)]the fact that copies of an arbitrary partially entangled pure state can be reduced from the Bell states by an LOCC transformation, where the ratio between the copies and the Bell states is inversely proportional to the entanglement entropy of the partially entangled state \cite{9}.
\end{description}

Extension of the above to multipartite states has been vigorously sought, but this albeit it is a hard problem.
Still, the multipartite results corresponding to (i) and (i$\hspace{-.1em}$i) have been given:
The tangle has been defined \cite{10}, which together with the concurrences gives a solution to (i) for three-qubit pure states. 
The tangle $\tau_{ABC}$ for three qubits $A$, $B$ and $C$ has an important property that $C^2_{A(BC)}=C^2_{AB}+C^2_{AC}+\tau_{ABC}$, where $C_{AB}$ is the concurrence between the qubits $A$ and $B$, $C_{AC}$ is the concurrence between the qubits $A$ and $C$, and $C_{A(BC)}$ is the concurrence between the qubit $A$ and the set of the qubits $B$ and $C$.

It has been shown that the entanglement of three-qubit pure states is expressed by five parameters \cite{12,13}.
With the coefficients of the generalized Schmidt decomposition, we can determine whether two three-qubit pure states are LU-equivalent or not \cite{18}.
The latter gives the result corresponding to (i$\hspace{-.1em}$i) for three-qubit pure states.
The result corresponding to (i$\hspace{-.1em}$i) for multipartite pure states, namely a necessary and sufficient condition for LU-equivalence, is given in Ref.\cite{Kraus}.

There are many other important researches to clarify the features of entanglement in multipartite systems.
In Ref. \cite{GHZW}, a stochastic LOCC classification of three-qubit pure states has been done.
This clarifies that there are six classes in the space of three-qubit pure states and gives a necessary and sufficient condition whether we can perform an LOCC transformation from a state to another state  with finite probability or not. 
The tangle can be expressed by the hyper-determinant \cite{11}.
A generalization of the Schmidt decomposition \cite{19} gives a sufficient condition for LU-equivalence of arbitrary multipartite pure states.
A set of operational entanglement measures which characterize the LU-equivalence classes of three-qubit pure states (up to complex conjugation) was recently given, and it was shown that we can determine operationally whether a state is LU-equivalent to its complex conjugate or not\cite{V}.
However, the results corresponding to (i$\hspace{-.1em}$i$\hspace{-.1em}$i)--(v$\hspace{-.1em}$i) have not been provided yet. 

In the present paper, we obtain the following four results.
First, a complete solution corresponding to (i$\hspace{-.1em}$i$\hspace{-.1em}$i) for three-qubit pure states is given. To be precise, we give an explicit necessary and sufficient condition to determine whether a deterministic LOCC transformation from an arbitrary three-qubit pure state $\left|\psi\right\rangle$ to another arbitrary state $\left|\psi'\right\rangle$ is executable or not.
We express the present condition in terms of the tangle, the concurrence between $A$ and $B$, the concurrence between $A$ and $C$, the concurrence between $B$ and $C$,  along with $J_{5}$, which is a kind of phase, and a new parameter $Q_{\mbox{e}}$, which means a kind of charge.
We thereby clarify the rules of conversion of the entanglement by arbitrary deterministic LOCC transformations.
Thus, defining the order between two states $\left|\psi'\right\rangle\preccurlyeq\left|\psi\right\rangle$ by the existence of an executable deterministic LOCC transformation from  $\left|\psi\right\rangle$ to $\left|\psi'\right\rangle$, we can make the whole set of three-qubit pure states a partially ordered set. 
To summarize the above, we find that three-qubit pure states are a partially ordered set parametrized by the six entanglement parameters.   
This is an extension of Nielsen's work \cite{20} to three-qubit pure states.

Second, as we already mentioned above, we introduce a new entanglement parameter $Q_{\mbox{e}}$.
The new parameter has the following three features:\\
1. Arbitrary three-qubit pure states are LU-equivalent if and only if their entanglement parameters $(C_{AB}, C_{AC}, C_{BC}, \tau_{ABC}, J_{5}, Q_{\mbox{e}})$ are the same.\\
2. The parameter $Q_{\mbox{e}}$ has a discrete value: $-1$, 0 or 1.\\
3. The complex conjugate transformation on $\left|\psi\right\rangle$ reverses the sign of $Q_{\mbox{e}}$.

Third, we clarify the rules of conversion of the entanglement by deterministic LOCC transformations.
We also find that we can interpret the conversion as the transfer and dissipation of the entanglement.

Fourth, we obtain the minimum number of times of measurements to reproduce an arbitrary deterministic LOCC transformation. The minimum number of times depends on the set of the initial and the final states of the deterministic LOCC transformation; we will list up the dependence in Table 1 in section 2. We also show that the order of measurements are commutable; we can choose which qubit is measured first, second and third.

After completing the present work, we noticed other important results \cite{22,32,24,25,26,27,28,29,30} which give partial solutions to (i$\hspace{-.1em}$i$\hspace{-.1em}$i) for three qubits.  
In particular, two recent studies \cite{25} and  \cite{30} are remarkable.
The former \cite{25} gives a necessary and sufficient condition of the possibility of a deterministic LOCC transformation of truly multipartite states whose tensor rank is two; the latter \cite{30} gives a necessary and sufficient condition of the possibility of a deterministic LOCC transformation of W-type states, both using approaches different from ours.
These studies \cite{25} and \cite{30} also give the results which correspond to the first column of Table 1. 
However, these studies have not achieved the complete solution to (i$\hspace{-.1em}$i$\hspace{-.1em}$i) for three-qubit pure states. 
Specifically, they cannot determine whether a deterministic LOCC transformation from an arbitrary GHZ-type truly tripartite state to an arbitrary bipartite state is possible or not.
Rules of conversion of entanglement have been provided only in implicit forms, and explicit forms of the rules have been yet to be given.
Nevertheless, we decided to employ these results \cite{25,30} partially in the proof of our main theory primarilly in order to shorten the proof.
Our original proof is given in the version 2 of this article in Arxiv.
We believe that our original proof still contains novel techniques and is valuable in its own right.

Let us overview the structure of the present paper.
In section 2, we present Theorems 1, 2 and 3 of the present paper, and overview their physical meanings.
Theorem 1 gives a necessary and sufficient condition of the LU-equivalence, Theorem 2 gives a necessary and sufficient condition of a deterministic LOCC transformation, and Theorem 3 gives the minimum
number of necessary times of measurements to reproduce an arbitrary deterministic LOCC transformation.
In section 3, we prove Theorem 1. 
In section 4, we prove Theorem 2.
In version 2 of this article in Arxiv, we prove Theorem 3. 

\section{Theorems and their physical meanings}
In this section, we list up Theorems in this article and describe their physical meaning. 

We consider only three-qubit pure states throughout the present paper.
Before listing up Theorems, we introduce a useful expression of three-qubit pure states.
An arbitrary pure state $\left|\psi\right\rangle$ of the three qubits $A$, $B$ and $C$ is expressed in the form of the generalized Schmidt decomposition
\begin{equation}
\left|\psi\right\rangle=\lambda_{0}\left|000\right\rangle+\lambda_{1}e^{i\varphi}\left|100\right\rangle+\lambda_{2}\left|101\right\rangle+\lambda_{3}\left|110\right\rangle+\lambda_{4}\left|111\right\rangle\label{L2.1}
\end{equation}
with a proper basis set \cite{18,60}. (There are two kinds of decompositions which are called  generalization of the Schmidt decomposition.
One was given in Ref. \cite{18} and the other was given in Ref. \cite{19}.
We use the former in the present paper.)
The coefficients $\{\lambda_{i}|i=0,...,4\}$ in \eqref{L2.1} are nonnegative real numbers and satisfy that $\sum^{4}_{i=0}\lambda^2_{i}=1$.
Note that the phase $\varphi$ can take any real values if one of the coefficients $\{\lambda_{i}|i=0,...,4\}$ is zero, in which case we define the phase $\varphi$ to be zero in order to remove the ambiguity.

Two different decompositions of the form \eqref{L2.1} are possible for the same state $\left|\psi\right\rangle$, one with $0\le\varphi\le\pi$ and the other with $\pi\le\varphi\le2\pi$.
These two decompositions are LU-equivalent; in other words, they can be transformed into each other by local unitary (LU) transformations.
Hereafter, we refer to the decomposition \eqref{L2.1} with $0\le\varphi\le\pi$ as the positive decomposition and the one \eqref{L2.1} with $\pi<\varphi<2\pi$ as the negative decomposition.
We also refer to the coefficients of the positive and negative decompositions as the positive-decomposition coefficients and the negative-decomposition coefficients, respectively.
Therefore, a set of coefficients gives a unique set of states that are LU-equivalent to each other, whereas such a set of states may give two possible sets of coefficients:
for $\varphi\ne0$, a set of positive-decomposition coefficients and a set of negative-decomposition coefficients are possible, while for $\varphi=0$, two sets of positive-decomposition coefficients are possible.
When $\sin\varphi\ne0$ holds, a set of LU-equivalent states and a set of positive-decomposition coefficients have a one-to-one correspondence.

We can express the entanglement parameters in the coefficients of the generalized Schmidt decomposition:
\begin{eqnarray}
C_{AB}&=&2\lambda_{0}\lambda_{3},\enskip C_{AC}=2\lambda_{0}\lambda_{2},\enskip C_{BC}=2|\label{concurrences}\lambda_{1}\lambda_{4}e^{i\varphi}-\lambda_{2}\lambda_{3}|,\\
\tau_{ABC}&=&4\lambda^2_{0}\lambda^2_{4}\label{tangle},\\
J_{5}&=&4\lambda^2_{0}(|\lambda_{1}\lambda_{4}e^{i\varphi}-\lambda_{2}\lambda_{3}|^2+\lambda^2_{2}\lambda^2_{3}-\lambda^2_{1}\lambda^2_{4})\label{J5},
\end{eqnarray}  
where $\tau$ is the tangle \cite{10}, $C_{AB}$, $C_{AC}$ and $C_{AC}$ are the concurrences \cite{6}, and $J_{5}$ is four times of $J_{5}$ in Ref. \cite{18}.

\begin{figure}
 \begin{minipage}{0.5\hsize}
  \begin{center}
   \includegraphics[width=70mm]{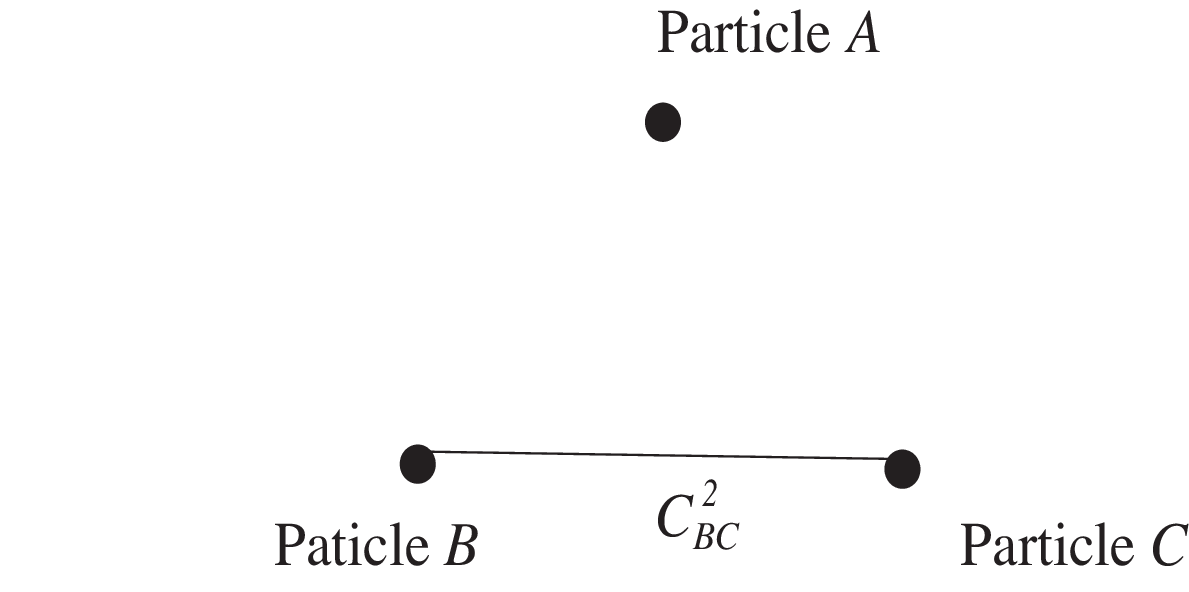}
   
   (a)
  \end{center}
 \end{minipage}
 \begin{minipage}{0.5\hsize}
  \begin{center}
   \includegraphics[width=70mm]{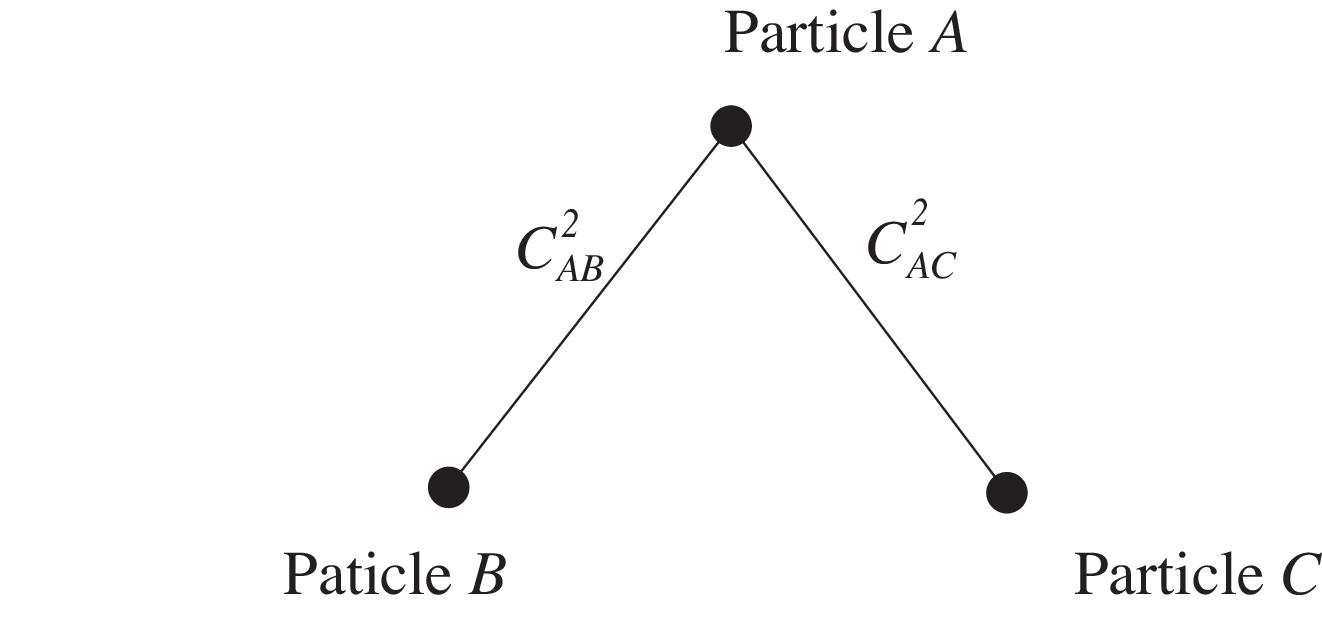}
   
   (b)
  \end{center}
 \end{minipage}
  \caption{The concept of (a) a biseparable state. There is no such state as (b).}
  \label{figfigfig}
\end{figure}

Finally, we define the names of types of states.
We refer to a state whose $\tau_{ABC}$ is nonzero or whose $C_{AB}$, $C_{AC}$ and $C_{BC}$ are all nonzero as a truly tripartite state.
We refer to a state which has only a single kind of the bipartite entanglement as a biseparable state (Fig.~\ref{figfigfig}(a)).
Note that there is no state which  has only two kinds of the bipartite entanglement (Fig.~\ref{figfigfig}(b)).
If there were such a state as in Fig.~\ref{figfigfig}(b), the coefficients $\{\lambda_{i},\varphi|i=0,...,4\}$ of the state would satisfy 
\begin{equation}
\lambda_{0}\lambda_{2}\ne0,\enskip\lambda_{0}\lambda_{3}\ne0,\enskip\lambda_{0}\lambda_{4}=0,\enskip|\lambda_{1}\lambda_{4}e^{i\varphi}-\lambda_{2}\lambda_{3}|=0,\label{hairi}
\end{equation}
but \eqref{hairi} is impossible.

The preparation has been now completed. Let us give Theorems and see their meaning. 
\begin{theorem}(Condition for the LU-equivalence)

Arbitrary three-qubit pure states are LU-equivalent if and only if their entanglement parameters $(C_{AB},C_{AC},C_{BC},\tau_{ABC},J_{5},Q_{\mbox{e}})$ are equal to each other. Here $Q_{\mbox{e}}$ is a new parameter which is given by 
\begin{equation}
Q_{\mbox{e}}=\mbox{sgn}\left[ \sin{\varphi}\left(\lambda^2_{0}-\frac{\tau_{ABC}+J_{5}}{2(C^2_{BC}+\tau_{ABC})}\right)\right],\label{Qe}
\end{equation}
where $\mbox{sgn}[x]$ is the sign function,
\begin{eqnarray}
\mbox{sgn}[x] =\left\{ \begin{array}{ll}
x/|x| & (x\ne0) \\
0 & (x=0) \\
\end{array} \right\},
\end{eqnarray}
and the parameters $\varphi$ and $\lambda_{0}$ are the coefficients of the generalized Schmidt decomposition.
\end{theorem}

\textbf{Physical meaning of Theorem 1:}
Theorem 1 gives a necessary and sufficient condition for the LU-equivalence of three-qubit pure states.
The necessary and sufficient conditions can be given in other ways.
However, the parameters used in Theorem 1 have very clear physical meaning of the magnitude, phase and charge of the entanglement, in the space of the LU-equivalence class of three-qubit pure states.

The concurrences $C_{AB}$, $C_{AC}$ and $C_{BC}$ express the amount of the entanglement between the qubits $A$ and $B$, $A$ and $C$ and $B$ and $C$, respectively.
The tangle $\tau_{ABC}$ expresses the amount of the entanglement among three qubits.

What about $J_{5}$?
The parameter $J_{5}$ and the concurrences let us derive a phase of the entanglement as follows:
\begin{equation}
\cos\varphi_{5}=\frac{J_{5}}{C_{AB}C_{AC}C_{BC}},\enskip0\le\varphi_{5}\le\pi.
\end{equation}
Let us refer to the phase $\varphi_{5}$ as the entanglement phase (EP).
The entanglement phase $\varphi_{5}$ is invariant with respect to local unitary operations, because all of the parameters $J_{5}$ and $C_{AB}$ $C_{AC}$ $C_{BC}$ are.
When $C_{AB}C_{AC}C_{BC}=0$, the phase becomes indefinite.
Hereafter, we refer to a state whose entanglement phase $\varphi_{5}$ is definite as an EP-definite state and to a state whose entanglement phase $\varphi_{5}$ is indefinite as an EP-indefinite state.
An EP-indefinite state with $\tau_{ABC}\ne0$ and an EP-definite state are truly tripartite states.
A truly tripartite state is an EP-indefinite state with $\tau_{ABC}\ne0$ or an EP-definite state.
A biseparable state is EP indefinite with $\tau_{ABC}=0$.
An EP-indefinite state with $\tau_{ABC}=0$ is a biseparable state.
 
Finally, we interpret the new parameter $Q_{\mbox{e}}$.
As we will prove in section 3, the parameter $Q_{\mbox{e}}$ is equal for the possible sets of coefficients of a state.
Therefore, the parameter $Q_{\mbox{e}}$ is invariant with respect to local unitary transformations.
The parameter $Q_{\mbox{e}}$ is a tripartite parameter, because $Q_{\mbox{e}}$ is invariant with respect to the permutation of the qubits $A$, $B$ and $C$. This fact is shown in Appendix A.
The complex-conjugate transformation of a state does not change the parameters $(C_{AB},C_{AC},C_{BC},\tau_{ABC},J_{5})$ nor $\lambda_{0}$, but reverses the sign of $\sin\varphi$. Thus, the complex-conjugate transformation reverses the sign of $Q_{\mbox{e}}$. As we have seen, the parameter $Q_{\mbox{e}}$ has characters that the electric charge has; hence, we refer to $Q_{\mbox{e}}$ as the entanglement charge.
\\

The next Theorem 2 gives a necessary and sufficient condition for the possibility of a deterministic LOCC.
In order to express Theorem 2 in simpler forms, we define  
three nonnegative real-valued parameters $K_{AB}$, $K_{AC}$ and $K_{BC}$ as follows:
\begin{eqnarray}
K_{AB}=C^2_{AB}+\tau_{ABC},\enskip K_{AC}=C^2_{AC}+\tau_{ABC},\enskip K_{BC}=C^2_{BC}+\tau_{ABC}.\label{KABKACKBC}
\end{eqnarray}
Then, the five parameters $K_{AB}$, $K_{AC}$, $K_{BC}$, $\tau_{ABC}$ and $J_{5}$ are independent of each other and are invariant with respect to local unitary operations. We can substitute these five parameters for the entanglement parameters $(C_{AB},C_{AC},C_{BC},\tau_{ABC},J_{5})$.
Let us refer to the old parameters $(C_{AB},C_{AC},C_{BC},\tau_{ABC},J_{5})$ as the $C$-parameters and to the new parameters $(K_{AB},K_{AC},K_{BC},\tau_{ABC},J_{5})$ as the $K$-parameters.
Note that $(C_{AB},C_{AC},C_{BC},\tau_{ABC},J_{5})$ and $(K_{AB},K_{AC},K_{BC},\tau_{ABC},J_{5})$ have a one-to-one correspondence.

We also define three parameters in order to simplify expressions which often appear in the present paper:
\begin{eqnarray}
J_{\mbox{ap}}\equiv C^2_{AB}C^2_{AC}C^2_{BC},\enskip
K_{\mbox{ap}}\equiv K_{AB}K_{AC}K_{BC},\enskip
K_{5}\equiv \tau_{ABC}+J_{5},\label{last1}\\
\Delta_{J}\equiv K^2_{5}-K_{\mbox{ap}}\ge 0,\label{Delta_{J}}
\end{eqnarray}
where the subscript ap is abbreviation of all pairs, and where $\Delta_{J}$ is sixteen times of $\Delta_{J}$ in Ref. \cite{18}. 
Note that these parameters $J_{\mbox{ap}}$, $K_{\mbox{ap}}$, $K_{5}$ and $\Delta_{J}$ are $not$ included in the $K$-parameters; we introduce them only for simplicity. 
By definition, $J_{\mbox{ap}}$, $K_{\mbox{ap}}$, $K_{5}$ and $\Delta_{J}$ are invariant with respect to local unitary transformations as well as permutations of $A$, $B$ and $C$.   

\begin{theorem}(deterministic LOCC)

A deterministic LOCC transformation from an arbitrary state $\left|\psi\right\rangle$ to another arbitrary state $\left|\psi'\right\rangle$ is executable if and only if the $K$-parameters of $\left|\psi\right\rangle$ and $\left|\psi'\right\rangle$ satisfy the following conditions:

Condition 1: There are real numbers $0\le\zeta_{A}\le1$, $0\le\zeta_{B}\le1$, $0\le\zeta_{C}\le1$ and $\zeta_{\mbox{lower}}\le\zeta\le1$ which satisfy the following equation:
\begin{equation}%\fl
\left(
\begin{array}{c}
K'_{AB} \\
K'_{AC} \\
K'_{BC} \\
\tau'_{ABC} \\
J'_{5}
\end{array}
\right)=\zeta
\left(
\begin{array}{ccccc}
\zeta_{A}\zeta_{B} &   &   &   &   \\
  & \zeta_{A}\zeta_{C} &  &   &   \\
  &   & \zeta_{B}\zeta_{C} &  &   \\
  &   &   & \zeta_{A}\zeta_{B}\zeta_{C} &   \\
  &   &   &   & \zeta_{A}\zeta_{B}\zeta_{C}
\end{array}
\right)\left(
\begin{array}{c}
K_{AB} \\
K_{AC} \\
K_{BC} \\
\tau_{ABC} \\
J_{5}
\end{array}
\right),\label{lastlast}
\end{equation}
where
\begin{equation}
\zeta_{\mbox{lower}}=\frac{J_{\mbox{ap}}}{(K_{AB}-\zeta_{C}\tau_{ABC})(K_{AC}-\zeta_{B}\tau_{ABC})(K_{BC}-\zeta_{A}\tau_{ABC})}.\label{nekolove}
\end{equation}

Condition 2: If the state $\left|\psi'\right\rangle$ is EP definite, we check whether 
\begin{equation}
(\Delta_{J}=0)\land(J_{\mbox{ap}}-J^2_{5}=0)\label{2012.1}
\end{equation}
holds or not.
When \eqref{2012.1} does not hold, the condition is
\begin{equation}
Q_{\mbox{e}}=Q'_{\mbox{e}}\enskip \mbox{and} \enskip\zeta=\tilde{\zeta},\label{revise12}
\end{equation}
where 
\begin{equation}%\fl
\tilde{\zeta}\equiv\frac{K_{\mbox{ap}}(J_{\mbox{ap}}-J^2_{5})+\Delta_{J}J_{\mbox{ap}}}{K_{\mbox{ap}}(J_{\mbox{ap}}-J^2_{5})+\Delta_{J}(K_{AB}-\zeta_{C}\tau_{ABC})(K_{AC}-\zeta_{B}\tau_{ABC})(K_{BC}-\zeta_{C}\tau_{ABC})}\label{8821}.
\end{equation}
When \eqref{2012.1} holds, the condition is 
\begin{equation}
|Q'_{\mbox{e}}|=\mbox{sgn}[(1-\zeta)(\zeta-\zeta_{\mbox{lower}})],\label{revise15}
\end{equation}
or in other expressions,
\begin{eqnarray}
Q'_{\mbox{e}}\left\{ \begin{array}{ll}
=0 & (\zeta=1\enskip\mbox{or}\enskip\zeta=\zeta_{\mbox{lower}}), \\
\ne0 & (otherwise). \\
\end{array} \right.
\end{eqnarray}
\end{theorem}

\textbf{Physical meaning of Theorem 2:}
Theorem 2 gives a necessary and sufficient condition for the possibility of a deterministic LOCC (d-LOCC) transformation.
Note that the conditions are given as the condition for a diagonal matrix which relates two vectors of the $K$-parameters.
This shows that there exists a vector structure in three-qubit pure states.

The difficulty of seeking necessary and sufficient conditions for the possibility of multipartite d-LOCC transformation lies in the fact that we cannot apply the majorization theory, which played an important part in clarifying bipartite-pure d-LOCC transformation, to multipartite pure states.
The majorization theory was applied to a vector structure which exists in the coefficients of the Schmidt decomposition.
Unfortunately, there is not a vector structure in the coefficients of the generalized Schmidt decomposition, but a tensor structure.
Thus, the majorization theory is not applicable to three-qubit pure states directly; this was the reason of the difficulty of clarification of three-qubit d-LOCC transformation.

However, Theorem 2 implies that there is a vector structure in three-qubit states in view of the entanglement measures.
Note that there is a possibility that other multipartite systems have similar structures;
it is plausible that the nonlocal features of $N$-qubit pure states can be expressed completely in the magnitudes, phases and charges of the entanglement.
Then, the approach of the present paper may be applicable to such systems. 

\begin{figure}
\begin{center}
\includegraphics[width=100mm]{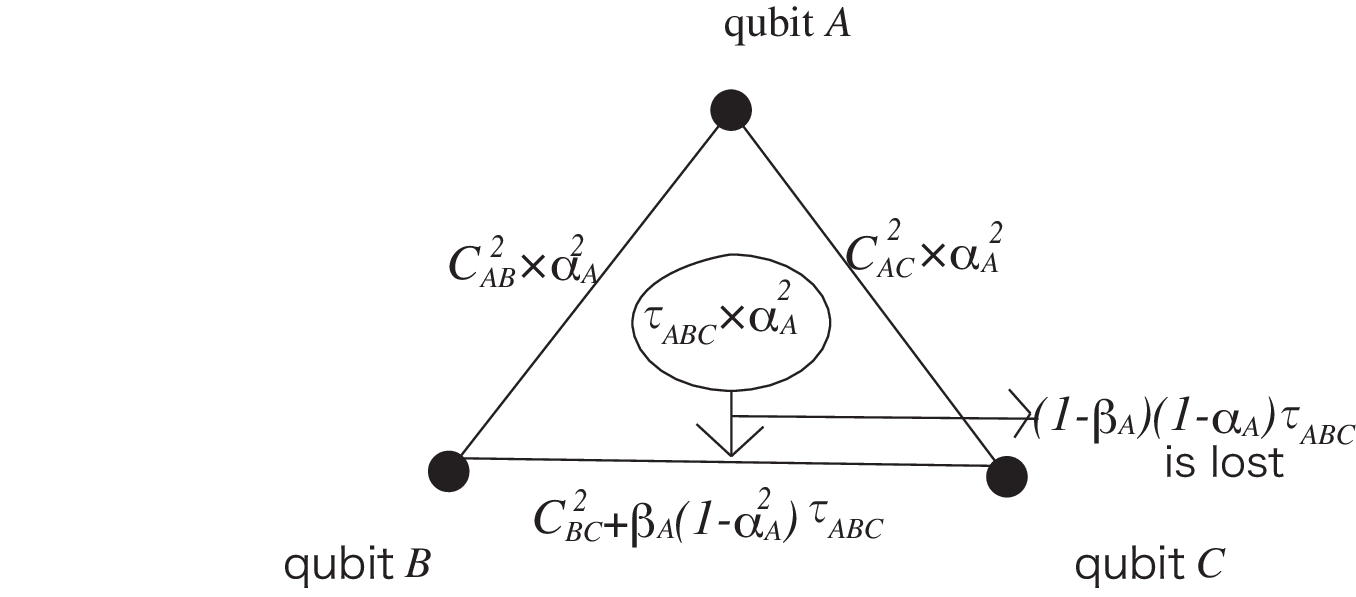}
\end{center}
\caption{Entanglement transfer}
\label{ETransfer}
\end{figure}

There are two important interpretations of Theorem 2.
One is the rule of the flow of the entanglement and the other is the preservation of the charge.
Let us see the rule of the flow of the entanglement.
Consider a d-LOCC transformation whose measurement is performed only on the qubit $A$.
In such a case, the condition 1 reduces to the following condition 1$'$: there are real numbers $0\le\zeta_{A}\le1$ and $\zeta_{\mbox{lower}}\le\zeta\le1$ which satisfy the following equation:
\begin{equation}%\fl
\left(
\begin{array}{c}
K'_{AB} \\
K'_{AC} \\
K'_{BC} \\
\tau'_{ABC} \\
J'_{5}
\end{array}
\right)=\zeta
\left(
\begin{array}{ccccc}
\zeta_{A} &   &   &   &   \\
  & \zeta_{A} &  &   &   \\
  &   & 1 &  &   \\
  &   &   & \zeta_{A} &   \\
  &   &   &   & \zeta_{A}
\end{array}
\right)\left(
\begin{array}{c}
K_{AB} \\
K_{AC} \\
K_{BC} \\
\tau_{ABC} \\
J_{5}
\end{array}
\right),\label{flowK}
\end{equation}
where
\begin{equation}
\zeta_{\mbox{lower}}=\frac{C^2_{BC}}{(K_{BC}-\zeta_{A}\tau_{ABC})}.
\end{equation}
Substituting  the $C$-parameters $(C_{AB},C_{AC},C_{BC},\tau_{ABC},J_{5})$  for the $K$-parameters in \eqref{flowK}, we obtain the following condition 1$''$:
$0\le\alpha_{A}\le1$, $0\le\beta_{A}\le1$ which satisfy the following equation: 
\begin{equation}
\left(
\begin{array}{c}
C'^2_{AB} \\
C'^2_{AC} \\
C'^2_{BC} \\
\tau'_{ABC} \\
J'_{5}
\end{array}
\right)=
\left(
\begin{array}{ccccc}
 \alpha_{A}^2 &   &   &   &   \\
  & \alpha_{A}^2 &  &   &   \\
  &   & 1 & \beta_{A}(1-\alpha_{A}^2) &   \\
  &   &   & \alpha_{A}^2 &   \\
  &   &   &   & \alpha_{A}^2
\end{array}
\right)\left(
\begin{array}{c}
C_{AB}^2 \\
C_{AC}^2 \\
C_{BC}^2 \\
\tau_{ABC} \\
J_{5}
\end{array}
\right).\label{flowJ}
\end{equation}
We can interpret the above as the rule how a deterministic measurement, which is a measurement whose results can be transformed into a unique state by local unitary operations without exception, changes the entanglement. We can express this change as in Fig.~\ref{ETransfer}.
After performing a deterministic measurement on the qubit $A$, the four entanglement parameters, $C^2_{AB}$, $C^2_{AC}$, $\tau_{ABC}$ and $J_{5}$, the last of which does not appear in Fig.~\ref{ETransfer}, are multiplied by $\alpha^2_{A}$.  
Note that these four entanglement parameters are related to the qubit $A$, which is the measured qubit.
The quantity $\beta_{A}(1-\alpha^2_{A})\tau_{ABC}$, which is a part of the entanglement lost from $\tau_{ABC}$, is added to $C^2_{BC}$, which is the only entanglement parameter that is not related to the measured qubit $A$. 
The quantity $(1-\beta_{A})(1-\alpha^2_{A})\tau_{ABC}$, which is the rest of the entanglement lost from $\tau_{ABC}$ disappear.
We call this phenomenon the entanglement transfer.

\begin{figure}
\begin{center}
\includegraphics[width=75mm]{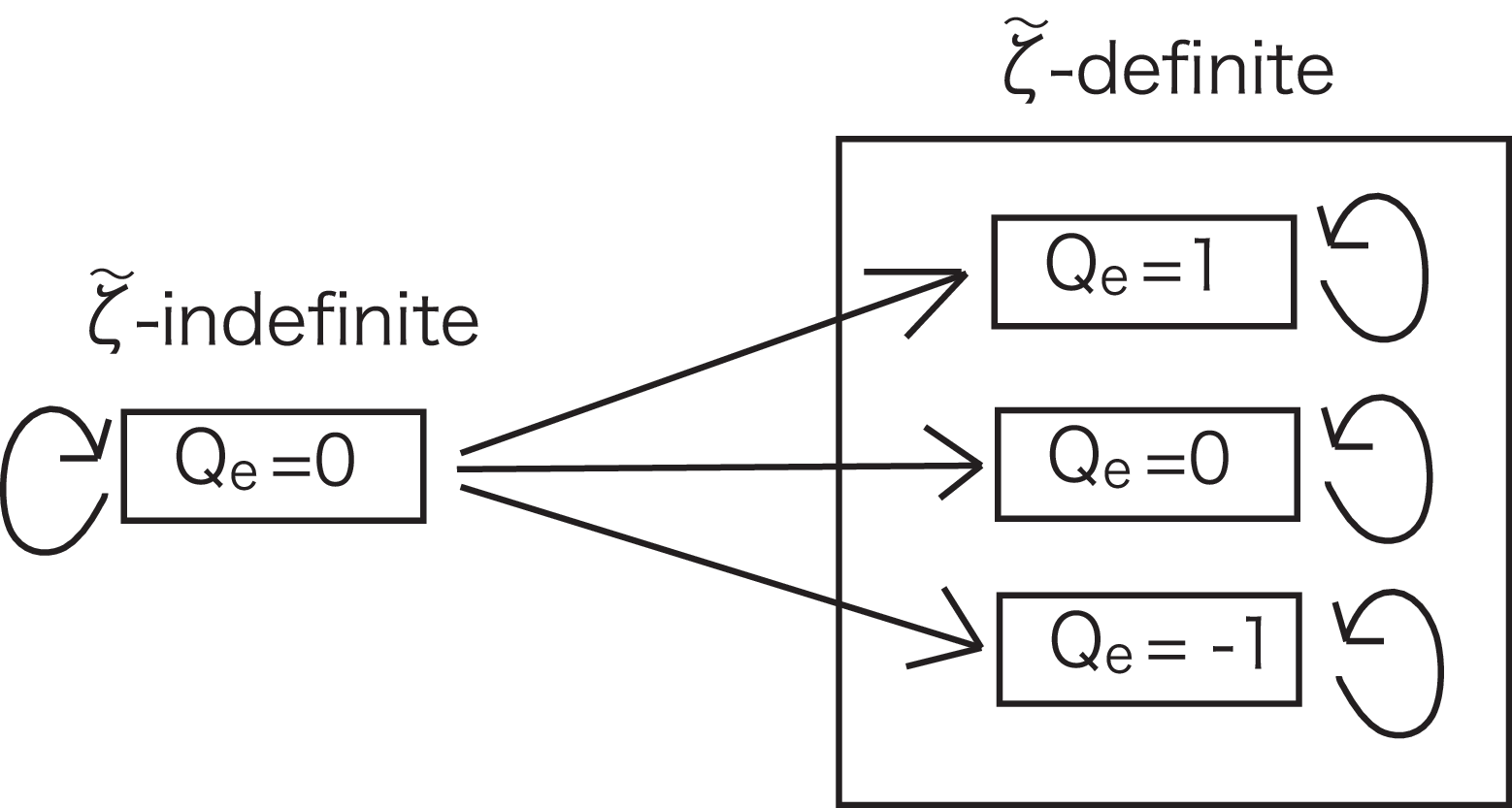}
\end{center}
\caption{The entanglement charge $Q_{\mbox{e}}$ for $\tilde{\zeta}$-definite states and $\tilde{\zeta}$-definite states. The arrows indicate the executable deterministic LOCC transformations among truly multipartite states; transformations not in this figure are not executable as deterministic LOCC transformations. For example, we cannot transform a $\tilde{\zeta}$-definite state whose $Q_{\mbox{e}}$ is 1 into another $\tilde{\zeta}$-definite state whose $Q_{\mbox{e}}$ is 0.}
 \label{denkakouzou}
\end{figure}

Finally, let us see the behavior of $Q_{\mbox{e}}$ when we perform a d-LOCC transformation.
Hereafter, we refer to a state which satisfies \eqref{2012.1} as $\tilde{\zeta}$-indefinite and refer to a state which does not satisfy \eqref{2012.1} as $\tilde{\zeta}$-definite.
The following statements hold:
\begin{description}
\item[Statement $\tilde{\zeta}$-1:]Any biseparable state is also a $\tilde{\zeta}$-indefinite state.
\item[Statement $\tilde{\zeta}$-2:]Any $\tilde{\zeta}$-indefinite state satisfies $Q_{\mbox{e}}=0$. 
\item[Statement $\tilde{\zeta}$-3:]A d-LOCC transformation from an EP-indefinite state to an EP-definite state is executable if and only if the initial state is $\tilde{\zeta}$-indefinite.
\item[Statement $\tilde{\zeta}$-4:]Among truly multipartite states, a d-LOCC transformation from a $\tilde{\zeta}$-indefinite state to a $\tilde{\zeta}$-definite state is executable, but the contrary is not executable.
\item[Statement $\tilde{\zeta}$-5:]When the initial state is $\tilde{\zeta}$-definite, the d-LOCC transformation conserves the entanglement charge $Q_{\mbox{e}}$.  
\end{description} 
Because of the above five statements, the $\tilde{\zeta}$-definite state can be considered as a ``charge-definite state.''
When we transform a $\tilde{\zeta}$-indefinite state into a $\tilde{\zeta}$-definite state, we can choose the value of the entanglement charge $Q_{\mbox{e}}$; once the value is determined, we cannot change it anymore with a deterministic LOCC transformation
 (Fig.~\ref{denkakouzou}). 
\\

The third theorem gives the minimum number of times of measurements to reproduce an arbitrary deterministic LOCC transformation between three-qubit pure states. 
\begin{theorem}{the minimum number of necessary times of measurements}
The minimum number of necessary times of measurements to reproduce an arbitrary deterministic
LOCC transformation from an arbitrary three-qubit state $\left|\psi\right\rangle$ to another arbitrary state $\left|\psi'\right\rangle$ is given as listed in Table 1.
\begin{table}
\caption{\label{times}The minimum number of times of measurements to reproduce an arbitrary deterministic LOCC transformation.}
\begin{center}
\begin{tabular}{@{}lll}
 \hline
Initial state &Final state&Times\\
\hline 
Truly tripartite state&Truly tripartite state&3\\
\hline
Truly tripartite state&Biseparable state or full-separable state&2\\
\hline
Biseparable state&Biseparable state or full-separable state&1\\
\hline
Full-separable state&Full-separable state&0\\
\hline
 
\end{tabular}
\end{center}
\end{table}
The order of measurements are commutable; we can choose which qubit is measured first, second and third.
\end{theorem}

\section{Proof of Theorem 1}
In this section, we prove Theorem 1.
We perform the proof in the following two steps.
First, we see that the $C$-paramters $(C_{AB}, C_{AC}, C_{BC}, \tau_{ABC}, J_{5})$ does not specify an LU-equivalence class uniquely. 
Second, we show that the entanglement charge $Q_{\mbox{e}}$ eliminate the non-uniqueness.

When we specify the set $(C_{AB}, C_{AC}, C_{BC}, \tau_{ABC}, J_{5})$, there are still two possible positive decompositions \cite{59,98}:
\begin{eqnarray}
(\lambda^{\pm}_{0})^2=\frac{J_{5}+\tau_{ABC}\pm\sqrt{\Delta_J}}{2(C^2_{BC}+\tau_{ABC})}=\frac{K_{5}\pm\sqrt{\Delta_J}}{2K_{BC}},  \label{ambi1}\\
(\lambda_{2}^\pm)^2=\frac{C^2_{AC}}{4(\lambda_0^\pm)^2},\enskip 
(\lambda_{3}^\pm)^2=\frac{C^2_{AB}}{4(\lambda_0^\pm)^2}, \enskip
(\lambda_{4}^\pm)^2=\frac{\tau_{ABC}}{4(\lambda_0^\pm)^2}, \label{ambi4}\\
(\lambda_{1}^\pm)^2=1-(\lambda_0^\pm)^2-\frac{C^2_{AB}+C^2_{AC}+\tau_{ABC}}{4(\lambda_0^\pm)^2},\label{ambi5}\\
\cos\varphi^\pm=\frac{(\lambda_1^\pm)^2(\lambda_4^\pm)^2+(\lambda_2^\pm)^2(\lambda_3^\pm)^2-C^2_{BC}/4}{2\lambda_1^\pm\lambda_2^\pm\lambda_3^\pm\lambda_4^\pm},\label{ambi6}
\end{eqnarray}
where
\begin{eqnarray}
0\le\varphi^{\pm}\le\pi.
\end{eqnarray}
Thus, there are four possible sets of coefficients for one set of $(C_{AB}, C_{AC}, C_{BC}, \tau_{ABC}, J_{5})$:
the positive-decomposition coefficients $\{\lambda^{+}_{i},\varphi^{+}|i=0,...,4\}$ and $\{\lambda^{-}_{i},\varphi^{-}|i=0,...,4\}$ as well as other two sets of coefficients $\{\lambda^{+}_{i},\tilde{\varphi}^{+}|i=0,...,4\}$ and $\{\lambda^{-}_{i},\tilde{\varphi}^{-}|i=0,...,4\}$, where $\tilde{\varphi}^{\pm}=2\pi-\varphi^{\pm}$ with $\pi\le\tilde{\varphi}^{\pm}\le2\pi$.
A state with $\{\lambda^{+}_{i},\varphi^{+}|i=0,...,4\}$ is LU-equivalent to a state with $\{\lambda^{-}_{i},\tilde{\varphi}^{-}|i=0,...,4\}$, while a state with $\{\lambda^{-}_{i},\varphi^{-}|i=0,...,4\}$ is LU-equivalent to a state with $\{\lambda^{+}_{i},\tilde{\varphi}^{+}|i=0,...,4\}$ \cite{59}. 
Therefore we can focus on two possible positive-decomposition coefficients $\{\lambda^{+}_{i},\varphi^{+}|i=0,...,4\}$ and $\{\lambda^{-}_{i},\varphi^{-}|i=0,...,4\}$ for a set of $(C_{AB}, C_{AC}, C_{BC}, \tau_{ABC}, J_{5})$.

Next, we prove that the entanglement charge $Q_{\mbox{e}}$ eliminates the non-uniqueness and that two states are LU-equivalent if and only if the six parameters $(C_{AB}, C_{AC}, C_{BC}, \tau_{ABC}, J_{5},Q_{\mbox{e}})$ of the two states are equal to each other.
If $Q_{\mbox{e}}\ne0$, we can determine one positive-decomposition coefficients and one negative-decomposition coefficients uniquely from the parameters $(C_{AB},C_{AC},C_{BC},\tau_{ABC},J_{5},Q_{\mbox{e}})$ as follows:
\begin{eqnarray}
\lambda_{0}^2=\frac{J_{5}+\tau_{ABC}\ddot{+}Q_{\mbox{e}}\sqrt{\Delta_J}}{2(C^2_{BC}+\tau_{ABC})}=\frac{K_{5}\ddot{+}Q_{\mbox{e}}\sqrt{\Delta_J}}{2K_{BC}},  \label{last3}\\
\lambda_{2}^2=\frac{C^2_{AC}}{4\lambda_0^2},\enskip
\lambda_{3}^2=\frac{C^2_{AB}}{4\lambda_0^2}, \enskip
\lambda_{4}^2=\frac{\tau_{ABC}}{4\lambda_0^2}, \label{last6}\\
\lambda_{1}^2=1-\lambda_0^2-\frac{C^2_{AB}+C^2_{AC}+\tau_{ABC}}{4\lambda_0^2},\label{last6.5}\\
\cos\varphi=\frac{\lambda_1^2\lambda_4^2+\lambda_2^2\lambda_3^2-C^2_{BC}/4}{2\lambda_1\lambda_2\lambda_3\lambda_4},\label{last7}
\end{eqnarray}
where $\ddot{+}$ is $+$ or $-$ when $\{\lambda_{i},\varphi|i=0,...,4\}$ are positive-decomposition coefficients or negative-decomposition coefficients, respectively.
Thus, if $Q_{\mbox{e}}\ne0$, the set of $(C_{AB},C_{AC},C_{BC},\tau_{ABC},J_{5})$ together with the entanglement 
charge $Q_{\mbox{e}}$ gives a unique set of LU-equivalent states.
Note that when $Q_{e}\ne0$, the parameter $Q_{\mbox{e}}$ is equal for the possible sets of coefficients of a state, because of \eqref{last3}--\eqref{last7}.

\begin{figure}
\begin{center}
\includegraphics[width=75mm]{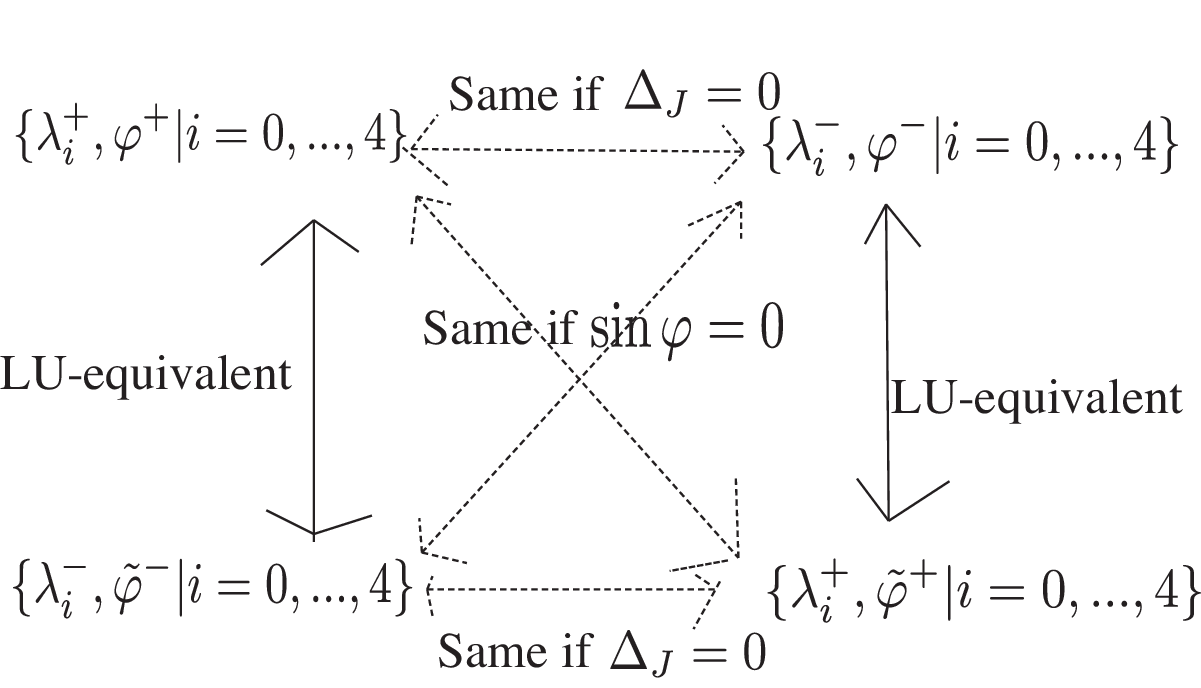}
\end{center}
\caption{The relation among the four sets of coefficients for $Q_{\mbox{e}}=0$. 
The relations indicated by solid lines are always valid, while those indicated by dotted lines are valid if and only if the noted conditions are satisfied.}
 \label{last8}
\end{figure}

If $Q_{\mbox{e}}=0$, at least one of $\sin\varphi$ and $\Delta_{J}$ is zero because of \eqref{Qe} and \eqref{ambi1}.
If $\sin\varphi$ is zero, $\{\lambda^{\pm}_{i},\varphi^{\pm}|i=0,...,4\}$ and $\{\lambda^{\pm}_{i},\tilde{\varphi}^{\pm}|i=0,...,4\}$ are equal.
If $\Delta_{J}$ is zero,  $\{\lambda^{+}_{i},\varphi^{+}|i=0,...,4\}$ and  $\{\lambda^{-}_{i},\varphi^{-}|i=0,...,4\}$ are equal as  $\{\lambda^{+}_{i},\tilde{\varphi}^{+}|i=0,...,4\}$ and  $\{\lambda^{-}_{i},\tilde{\varphi}^{-}|i=0,...,4\}$ are, respectively, because of \eqref{ambi1}.
Thus, if $Q_{\mbox{e}}$ is zero, the four sets of coefficients $\{\lambda^{+}_{i},\varphi^{+}|i=0,...,4\}$, $\{\lambda^{-}_{i},\varphi^{-}|i=0,...,4\}$, $\{\lambda^{+}_{i},\tilde{\varphi}^{-}|i=0,...,4\}$ and $\{\lambda^{-}_{i},\tilde{\varphi}^{-}|i=0,...,4\}$ are LU-equivalent (Fig.~\ref{last8}).
Thus, if $Q_{\mbox{e}}=0$, the set of $(C_{AB},C_{AC},C_{BC},\tau_{ABC},J_{5})$ gives a unique set of LU-equivalent states.
Note that the above guarantees that when $Q_{\mbox{e}}=0$, the parameter $Q_{\mbox{e}}$ is equal for the possible sets of coefficients of a state.
Incidentally, a state is LU-equivalent to its complex-conjugate if and only if its entanglement charge $Q_{\mbox{e}}$ is zero. The complex conjugate transformation of the state only changes the sign of $\varphi$. Thus, a state is LU-equivalent to its complex conjugate if and only if $\{\lambda^{\pm}_{i},\varphi^{\pm}|i=0,...,4\}$ are LU-equivalent to $\{\lambda^{\pm}_{i},\tilde{\varphi}^{\pm}|i=0,...,4\}$; this LU-equivalence is illustrated in Fig.~\ref{last8}.

For the reasons stated above, the set of $(C_{AB},C_{AC},C_{BC},\tau_{ABC},J_{5})$ together with the entanglement 
charge $Q_{\mbox{e}}$ gives a unique set of LU-equivalent states.
In other words, two states are LU-equivalent if and only if the $C$-parameters $(C_{AB},C_{AC},C_{BC},\tau_{ABC},J_{5},Q_{\mbox{e}})$ of the two states are equal to each other.
This is the statement of Theorem 1, and thus the proof is completed.
$\Box$

Note that a state is LU-equivalent to its complex conjugate if and only if $Q_{\mbox{e}}=0$.
Thus, the condition $Q_{\mbox{e}}=0$ is equivalent to $E_{6}=0$ in Ref. \cite{V}.

\section{The Proof of Theorem 2}
In this section, we prove Theorem 2.
All the deterministic LOCC transformations are categorized into any of the following cases determined by the initial and final states:
\begin{description}
\item[Case $\mathfrak{A}$:]Both of the initial and final states are EP definite and the initial state is GHZ-type.
\item[Case $\mathfrak{B}$:]The initial state is EP definite and the final state is EP indefinite.
\item[Case $\mathfrak{C}$:]The initial state is EP indefinite and GHZ-type.
\item[Case $\mathfrak{D}$:]The tangle of initial state is zero.
\end{description}
Note that these four Cases exhaust all cases of the initial and final states; 
$(Proof$: The case where both the initial and final states are EP-definite is exhausted by Cases $\mathfrak{A}$, $\mathfrak{C}$ and $\mathfrak{D}$.
The case where the initial state is EP-definite and the final state is EP-indefinite is Case $\mathfrak{B}$.
The case where the initial state is EP-indefinite is exhausted by Cases $\mathfrak{C}$ and $\mathfrak{D}$.$\Box)$

We carry out the proof of each Case in sections 4.1--4.4, respectively.

\subsection{Case $\mathfrak{A}$}
First, we prove Theorem 2 in Case $\mathfrak{A}$, where both of the initial and final states are EP definite and the initial state is GHZ-type.
We will argue that we can assume the final state to be GHZ-type too.
The final state after an LOCC transformation is EP definite, and hence the final state is a truly multipartite state.
We cannot reach a W-type state from a GHZ-type initial state with an LOCC transformation \cite{GHZW}.
Thus the final state must be GHZ-type.
We can also show that if the initial and final states satisfy the two conditions of Theorem 2, the final state must be GHZ-type.
$(Proof$:  If the final state would not be GHZ-type, then $\tau'_{ABC}$ would be zero, and thus at least one of $\zeta$, $\zeta_{A}$, $\zeta_{B}$ and $\zeta_{C}$ would be zero.
Then, at least one of $K'_{AB}$, $K'_{AC}$ and $K'_{BC}$ would be zero.
Because of $K'_{\mbox{ap}}\ge J'_{\mbox{ap}}$, the final state would not be W-type.
An EP-definite state is GHZ-type or W-type, and thus this is a contradiction. $\Box)$
Thus, in the present case, we only have to consider the case in which both initial and final states are GHZ-type.
In this case, a necessary and sufficient condition of the possibility of a d-LOCC transformation is already given \cite{25}.
Thus, in the present case, we only have to prove the equivalence between the conditions of Theorem 2 and the necessary and sufficient condition. 

An arbitrary GHZ-type state can be expressed as follows \cite{25}:
\begin{equation}   
\left|\psi\right\rangle=\frac{1}{\sqrt{N}}(\left|\tilde{0}_{A}\tilde{0}_{B}\tilde{0}_{C}\right\rangle+z\left|\tilde{1}_{A}\tilde{1}_{B}\tilde{1}_{C}\right\rangle),\label{cdef}
\end{equation}
where $\{\left|\tilde{0}_{i}\right\rangle|i=A,B,C\}$ and $\{\left|\tilde{1}_{i}\right\rangle|i=A,B,C\}$ are normalized states of the qubits $A$, $B$ and $C$, with their relative phases adjusted such that all of $c_{i}\equiv \left\langle\tilde{0}_{i}|\tilde{1}_{i}\right\rangle$ are real and non-negative, while  $z$ is an arbitrary complex number and $N$ is the normalization constant.
There are two possible expressions of the form \eqref{cdef} for an arbitrary state.
These two expressions have the same values of $\{c_{i}\}$ but different values of $z$.
The relation between the two values of $z$ is as follows:
\begin{eqnarray}
\mbox{If there is no zeros in the set $\{c_{i}\}$}: \enskip z_{1}=\frac{1}{z_{2}},\label{zzother1}\\
\mbox{If there is a zero in the set $\{c_{i}\}$}: \enskip |z_{1}|=\frac{1}{|z_{2}|},\label{zzother2}
\end{eqnarray}
where $z_{1}$ and $z_{2}$ are the values of $z$ in the two possible expressions of the form \eqref{cdef}.

When the set $\{c_{i}\}$ of the states $\left|\psi\right\rangle$ and the set $\{c'_{i}\}$ of $\left|\psi'\right\rangle$ have no zeros, a d-LOCC transformation from the state $\left|\psi\right\rangle$ to the state $\left|\psi'\right\rangle$ is executable if and only if the following expressions are satisfied \cite{25}:
\begin{eqnarray}
c'_{i}\ge c_{i}\enskip (k=A,B,C)\label{revise6},\\
\frac{s'}{s}=\frac{c_{A}c_{B}c_{C}}{c'_{A}c'_{B}c'_{C}}\label{revise7},\\
\frac{n'}{n}=\frac{c_{A}c_{B}c_{C}}{c'_{A}c'_{B}c'_{C}}\label{revise8},
\end{eqnarray}
where
\begin{eqnarray}
n\equiv\frac{2\mbox{Re}[z]}{|z|^2+1}, \enskip s\equiv\frac{2\mbox{Im}[z]}{|z|^2-1}\label{nands}
\end{eqnarray}
with the parameter $z$ of the state $\left|\psi\right\rangle$, and $s'$ and $n'$ are defined for the  state $\left|\psi'\right\rangle$ with the parameter $z'$.
The parameter $z$ can take the two different values, but $n$ and $s$ have the same values for each value of $z$, except for $z=\pm1$; when $(|z|=1)\land(z\ne\pm1)$, we consider the parameter $s$ to take one value $\infty$.
In the case where $z=\pm1$, the parameter $s$ becomes indefinite; it becomes $0/0$.
When the parameter $s$ becomes indefinite, we only leave out \eqref{revise7}.
The parameter $s$ is indefinite if and only if $z=\pm1$.
As we will show later, the equation $z=\pm1$ is equivalent to \eqref{2012.1} in Theorem 2.
There are two other cases where we have to treat \eqref{revise7} and \eqref{revise8} exceptionally;
the case where at least one of $s$ and $s'$ becomes $0$ or $\infty$, and the case where at least one of $n$ and $n'$ becomes $0$.
In the former case, we consider that \eqref{revise7} holds if and only if $s$ and $s'$ has the same value; for example, if $s$ takes $\infty$, the equation \eqref{revise7} holds if and only if $s'=\infty$.
In the same manner, in the latter case, we consider that \eqref{revise8} holds if and only if $n$ and $n'$ has the same value.

Let us prove that \eqref{revise6}--\eqref{revise8} are equivalent to the conditions of Theorem 2.
We can express the paramters $z$ and $\{c_{i}\}$ in terms of the $C$-paramters:
\begin{eqnarray}
z&=&-\frac{\sqrt{K_{AB}K_{AC}}}{2\lambda^2_{0}\sqrt{K_{BC}}}e^{-i\tilde{\varphi}_{5}},\label{revise11}\\
c_{A}&=&\frac{C_{BC}}{\sqrt{K_{BC}}},\enskip c_{B}=\frac{C_{AC}}{\sqrt{K_{AC}}},\enskip c_{C}=\frac{C_{AB}}{\sqrt{K_{AB}}},\label{revise4}
\end{eqnarray}
where 
\begin{equation}
e^{-i\tilde{\varphi}_{5}}=\frac{\lambda_{2}\lambda_{3}-\lambda_{1}\lambda_{4}e^{i\varphi}}{|\lambda_{2}\lambda_{3}-\lambda_{1}\lambda_{4}e^{i\varphi}|}=\cos\varphi_{5}-i\mbox{sgn}[\sin\varphi]\sin\varphi_{5}.\label{deftilvarphi}
\end{equation}
Because there are two possible sets of $\{\lambda_{i}, \varphi|i=0,...,4\}$ in \eqref{revise11}, the parameter $z$ takes two values, which are $z_{1}$ and $z_{2}$ in \eqref{zzother1} and \eqref{zzother2}.
The derivation of the above expressions is given in Appendix B.
It is easily seen from \eqref{revise4} that the set $\{c_{i}\}$ of a state has no zeros if and only if the state is EP-definite.
Now that we have reduced Case $\mathfrak{A}$ into the case in which the initial and final states are EP-definite and GHZ-type, we only have to prove that \eqref{revise6}--\eqref{revise8} are equivalent to the conditions of Theorem 2. 

Using \eqref{revise11} and \eqref{revise4}, we prove that \eqref{revise6} and \eqref{revise8} are equivalent to the condition 1 of Theorem 2 except for $\zeta_{\mbox{lower}}\le\zeta\le1$.
Substituting \eqref{revise4} into \eqref{revise6}, we transform \eqref{revise6} into
\begin{eqnarray}
\frac{\tau'_{ABC}}{\tau_{ABC}}\le\frac{K'_{AB}}{K_{AB}},\enskip\frac{\tau'_{ABC}}{\tau_{ABC}}&\le&\frac{K'_{AC}}{K_{AC}},\enskip\frac{\tau'_{ABC}}{\tau_{ABC}}\le\frac{K'_{BC}}{K_{BC}}.\label{revise9}
\end{eqnarray}
Because of \eqref{deftilvarphi},
\begin{equation}
\mbox{Re}[z]=-\frac{\sqrt{K_{AB}K_{AC}}}{2\lambda^2_{0}\sqrt{K_{BC}}}\mbox{cos}\varphi_{5}.
\end{equation}
Because the possible two sets of $\{\lambda_{i},\varphi|i=0,...,1\}$ are $\{\lambda^{+}_{i},\varphi^{+}|i=0,...,1\}$ 
and $\{\lambda^{-}_{i},\tilde{\varphi}^{-}|i=0,...,1\}$ or
 $\{\lambda^{-}_{i},\varphi^{-}|i=0,...,1\}$ and 
 $\{\lambda^{+}_{i},\tilde{\varphi}^{+}|i=0,...,1\}$,
we obtain
\begin{equation}
n=-\frac{\sqrt{K_{\mbox{ap}}}}{2K_{5}}\mbox{cos}\varphi_{5}.
\end{equation}
Thus, we can trasform \eqref{revise8} into
\begin{eqnarray}
\frac{K'_{5}}{K_{5}}=\frac{J'_{5}}{J_{5}}\enskip(\mbox{for }n\ne0),\enskip\enskip J_{5}=J'_{5}=0\enskip(\mbox{for }n=0).\label{revise10}
\end{eqnarray}
Let us then prove that the expressions \eqref{revise9} and \eqref{revise10} are equivalent to the existence of $0\le\zeta\le1$, $0\le\zeta_{A}\le1$, $0\le\zeta_{B}\le1$ and $0\le\zeta_{C}\le1$, which satisfy \eqref{lastlast}.
To show this, we only have to define $\zeta$, $\zeta_{A}$, $\zeta_{B}$ and $\zeta_{C}$ as follows:
\begin{eqnarray}
\zeta_{A}=\frac{K_{BC}\tau'_{ABC}}{K'_{BC}\tau_{ABC}},\enskip\zeta_{B}=\frac{K_{AC}\tau'_{ABC}}{K'_{AC}\tau_{ABC}},\enskip\zeta_{C}=\frac{K_{ABf}\tau'_{ABC}}{K'_{AB}\tau_{ABC}},\label{revise23}\\
\zeta=\frac{\tau'_{ABC}}{\tau_{ABC}\zeta_{A}\zeta_{B}\zeta_{C}}.\label{revise24}
\end{eqnarray}
Because of \eqref{revise10}, \eqref{revise23} and \eqref{revise24}, the parameters $\zeta$, $\zeta_{A}$, $\zeta_{B}$ and $\zeta_{C}$ satisfy \eqref{lastlast}.
Because of \eqref{revise9}, the parameters $\zeta_{A}$, $\zeta_{B}$ and $\zeta_{C}$ are less than or equal to one.
Because of \eqref{revise23} and \eqref{revise24}, the parameters $\zeta_{A}$, $\zeta_{B}$ and $\zeta_{C}$ are greater than or equal to zero.
Thus, the expressions \eqref{revise6} and \eqref{revise8} are equivalent to the condition 1 of Theorem 2, except for $\zeta_{\mbox{lower}}\le\zeta\le1$.

Now we have proven that the equations \eqref{revise6} and \eqref{revise8}, which are the ones on the parameters $\{c_{i}\}$ and $n$, are equivalent to the condition 1 of Theorem 2 except for $\zeta_{\mbox{lower}}\le\zeta\le1$.
Next, let us prove that the condition on the parameter $s$ is equivalent to the inequality $\zeta_{\mbox{lower}}\le\zeta\le1$ and the condition 2 of Theorem 2.
Hereafter, we assume the condition 1 of Theorem 2 except for $\zeta_{\mbox{lower}}\le\zeta\le1$ until the end of Case $\mathfrak{A}$.
The condition on $s$ is \eqref{revise7} for $z\ne\pm1$ but is left out for $z=\pm1$.
First, we prove that the equation $z=\pm1$ is equivalent to \eqref{2012.1}.
Because of \eqref{revise11}, we can transform $z=\pm1$ into 
\begin{eqnarray}
z=\pm1&\Leftrightarrow&(\tilde{\varphi}_{5}=0,\pi)\land\left(\frac{\sqrt{K_{AB}K_{AC}}}{2\lambda^2_{0}\sqrt{K_{BC}}}=1\right)\\
&\Leftrightarrow&(J_{\mbox{ap}}\sin^2\varphi_{5}=0)\land\left(\frac{\sqrt{K_{\mbox{ap}}}}{K_{5}\pm \sqrt{\Delta_{J}}}=1\right)\label{accept1}\\
&\Leftrightarrow&(J_{\mbox{ap}}\sin^2\varphi_{5}=0)\land(\Delta_{J}=0)\\
&\Leftrightarrow&(J_{\mbox{ap}}-J^2_{5}=0)\land(\Delta_{J}=0).
\end{eqnarray}
where the double sign $\pm$ in \eqref{accept1} is the one in \eqref{ambi1}; note that the possible two sets of $\{\lambda_{i},\varphi|i=0,...,1\}$ are $\{\lambda^{+}_{i},\varphi^{+}|i=0,...,1\}$ 
and $\{\lambda^{-}_{i},\tilde{\varphi}^{-}|i=0,...,1\}$ or
 $\{\lambda^{-}_{i},\varphi^{-}|i=0,...,1\}$ and 
 $\{\lambda^{+}_{i},\tilde{\varphi}^{+}|i=0,...,1\}$.

Thus, the equation $z=\pm1$ is equivalent to \eqref{2012.1}.
In other words, Eq. \eqref{2012.1} does not hold for $z\ne\pm1$.

Let us next prove that the condition on $s$ is equivalent to the condition 2 of Theorem 2 and $\zeta_{\mbox{lower}}\le\zeta\le1$ in the case $z\ne\pm1$.
In this case, what we have to prove is that \eqref{revise7} is equivalent to the condition 2 of Theorem 2 and $\zeta_{\mbox{lower}}\le\zeta\le1$.
In this case, \eqref{2012.1} does not hold, and thus $\left|\psi\right\rangle$ is $\tilde{\zeta}$-definite.
Hence, we only have to prove that \eqref{revise7} is equivalent to \eqref{revise12}.
(Note that $\zeta_{\mbox{lower}}\le\tilde{\zeta}\le1$.)
To prove the equivalence of \eqref{revise7} and \eqref{revise12}, we first express $s$ in terms of the $K$-parameters.
When $Q_{\mbox{e}}\ne0$, by substituting \eqref{last3} and \eqref{deftilvarphi} into \eqref{revise7}, we obtain
\begin{equation}
s=-\frac{\sqrt{K_{\mbox{ap}}}}{Q_{\mbox{e}}\sqrt{\Delta_{J}}}\sin\varphi_{5}.\label{ver4-5}
\end{equation}
Because of \eqref{revise4} and \eqref{ver4-5}, 
\begin{equation}
\frac{Q'_{\mbox{e}}\sqrt{\Delta_{J}}}{Q_{\mbox{e}}\sqrt{\Delta'_{J}}}\frac{\sqrt{J'_{\mbox{ap}}}\sin\varphi'_{5}}{\sqrt{J_{\mbox{ap}}}\sin
\varphi_{5}}=1\label{revise14}
\end{equation}
always holds. Substituting $\Delta_{J}=K^2_{5}-K_{\mbox{ap}}$, $J_{\mbox{ap}}\sin^2\varphi_{5}=J_{\mbox{ap}}-J^2_{5}$ and \eqref{lastlast} into \eqref{revise14} and reducing it, we can obtain $Q_{\mbox{e}}=Q'_{\mbox{e}}$ and $\zeta=\tilde{\zeta}$.
Thus, when $Q_{\mbox{e}}\ne0$, \eqref{revise12} is equivalent to \eqref{revise7}.

Now we consider the case where $z\ne\pm1$, and thus when $Q_{\mbox{e}}=0$, only one of the expressions $\mbox{sin}\varphi_{5}=0$ and $\Delta_{J}=0$ holds.
When $\mbox{sin}\varphi_{5}=0$ holds, the equation \eqref{revise7} is equivalent to $s=s'=0$.
When $\Delta_{J}=0$ holds, the equation \eqref{revise7} is equivalent to $s=s'=\infty$.
Because of \eqref{nands} and \eqref{revise11}, the equations $s=s'=0$ and $s=s'=\infty$ are equivalent to $\mbox{sin}\varphi_{5}=\mbox{sin}\varphi'_{5}=0$ and $\Delta_{J}=\Delta'_{J}=0$, respectively.
Because of $(Q_{\mbox{e}}=0)\Leftrightarrow((\mbox{sin}\varphi_{5}=0)\lor(\Delta_{J}=0))$, $\Delta_{J}=K^2_{5}-K_{\mbox{ap}}$, $J_{\mbox{ap}}\sin^2\varphi_{5}=J_{\mbox{ap}}-J^2_{5}$ and \eqref{lastlast}, when the equation $\mbox{sin}\varphi_{5}=0$ holds, the equation $\mbox{sin}\varphi_{5}=\mbox{sin}\varphi'_{5}=0$ is equivalent to \eqref{revise12}.
In the same manner, when the equation $\Delta_{J}=0$ holds, $\Delta_{J}=\Delta'_{J}=0$ is equivalent to \eqref{revise12}.
Thus, when $Q_{\mbox{e}}=0$, the expression \eqref{revise12} is equivalent to \eqref{revise7}.
We have already proven that when $Q_{\mbox{e}}\ne0$, the expression \eqref{revise12} is equivalent to \eqref{revise7}.
Thus, \eqref{revise12} is equivalent to \eqref{revise7}.

Finally, we consider the case $z=\pm1$.
In this case, there is no condition of $s$ and the condition $\eqref{2012.1}$ holds.
Because of \eqref{2012.1}, the condition 2 of Theorem 2 becomes \eqref{revise15} in this case.
Thus, we only have to prove that \eqref{2012.1} is equivalent to $\zeta_{\mbox{lower}}\le\zeta\le1$ and \eqref{revise15}.
Because we have assumed the condition 1 of Theorem 1 except for $\zeta_{\mbox{lower}}\le\zeta\le1$, we obtain the following equations:
\begin{eqnarray}
\frac{J'^2_{5}}{J'_{\mbox{ap}}}&=&\frac{\zeta_{\mbox{lower}}}{\zeta}\frac{J^2_{5}}{J_{\mbox{ap}}},\label{ver4-3}\\
\Delta'_{J}&=&(\zeta\zeta_{A}\zeta_{B}\zeta_{C})^2(K^2_{5}-\zeta K_{\mbox{ap}})\ge(\zeta\zeta_{A}\zeta_{B}\zeta_{C})^2\Delta_{J}.\label{ver4-4}
\end{eqnarray}
The derivation of \eqref{ver4-3} and \eqref{ver4-4} is as follows:
\begin{eqnarray}
\frac{J'^2_{5}}{J'_{\mbox{ap}}}&=&\frac{(\zeta\zeta_{A}\zeta_{B}\zeta_{C})^2J^2_{5}}{\zeta^3\zeta^2_{A}\zeta^2_{B}\zeta^2_{C}(K_{BC}-\zeta_{A}\tau_{ABC})(K_{AC}-\zeta_{B}\tau_{ABC})(K_{AB}-\zeta_{C}\tau_{ABC})}\nonumber\\
&=&\frac{\zeta_{\mbox{lower}}}{\zeta}\frac{J^2_{5}}{J_{\mbox{ap}}},\\
\Delta'_{J}&=&(\zeta\zeta_{A}\zeta_{B}\zeta_{C})^2K^2_{5}-\zeta^3\zeta^2_{A}\zeta^2_{B}\zeta^2_{C}K_{\mbox{ap}}\nonumber\\
&=&(\zeta\zeta_{A}\zeta_{B}\zeta_{C})^2(K^2_{5}-\zeta K_{\mbox{ap}})\ge(\zeta\zeta_{A}\zeta_{B}\zeta_{C})^2\Delta_{J}.
\end{eqnarray}

By using \eqref{ver4-3} and \eqref{ver4-4}, let us prove that \eqref{2012.1} is equivalent to $\zeta_{\mbox{lower}}\le\zeta\le1$ and \eqref{revise15}.
First, we prove that \eqref{2012.1} is a sufficient condition of $\zeta_{\mbox{lower}}\le\zeta\le1$ and \eqref{revise15}
When \eqref{2012.1} is valid, the equation $J^2_{5}/J_{\mbox{ap}}=1$ holds.
Because $J'^2_{5}/J'_{\mbox{ap}}=\cos^2\varphi'_{5}\le1$ is always valid, when \eqref{2012.1} is valid, the inequality $\zeta_{\mbox{lower}}\le\zeta$ follows \eqref{ver4-3}.
In the same manner, when \eqref{2012.1} is valid, the inequality $\zeta\le1$ follows \eqref{ver4-4}. 
Then, note that the equation $J'^2_{5}=J'_{\mbox{ap}}$ holds if and only if $\zeta=\zeta_{\mbox{lower}}$ holds, and that the equation $\Delta'_{J}=0$ holds if and only if  $\zeta=1$ holds.
Because of \eqref{Qe}, the equation $Q'_{\mbox{e}}=0$ holds if and only if $(\Delta'_{J}=0)\lor(J'^2_{5}=J'_{\mbox{ap}})$ holds.
Thus, when \eqref{2012.1} is valid, $Q'_{\mbox{e}}=0$ if and only if $\zeta=\zeta_{\mbox{lower}}$ or $\zeta=1$.
In the other words, now we have proven that \eqref{2012.1} is a sufficient condition of $\zeta_{\mbox{lower}}\le\zeta\le1$ and \eqref{revise15}.

Let us prove that \eqref{2012.1} is also a necessary condition.
Because of  $\zeta_{\mbox{lower}}\le\zeta\le1$, \eqref{ver4-3}, \eqref{ver4-4} and \eqref{Qe}, the equation $|Q'_{\mbox{e}}|=\mbox{sgn}[(1-\zeta)(\zeta-\zeta_{\mbox{lower}})]$ is valid if and only if both of $\Delta_{J}$ and $J^2_{5}-J_{\mbox{ap}}$ are zero.
In the other words, \eqref{2012.1} is also a necessary condition of $\zeta_{\mbox{lower}}\le\zeta\le1$ and \eqref{revise15}.
Thus, \eqref{2012.1} is equivalent to  $\zeta_{\mbox{lower}}\le\zeta\le1$ and \eqref{revise15}, and thus we have completed the proof in Case $\mathfrak{A}$.
$\Box$

\subsection{Case $\frak{B}$}
In this subsection, we prove Theorem 2 in Case $\mathfrak{B}$, where the initial state is EP-definite and the final state is EP-indefinite.
In the proof, we will use the following two lemmas which describe how a measurement changes entanglement.
\begin{lemma}
Let us consider the situation where a measurement $\{M_{(i)}\}$ is performed on the qubit $A$ of an arbitrary three-qubit state $\left|\psi\right\rangle$.
The state $\left|\psi^{(i)}\right\rangle\equiv M_{(i)}\left|\psi\right\rangle/\sqrt{p_{(i)}}$ is obtained as the $i$-th result with the probability $p_{(i)}$.
Then, the following equations are valid: 
\begin{eqnarray}
C^{(i)}_{AB}=\alpha^{(i)}C_{AB},\enskip
C^{(i)}_{AC}=\alpha^{(i)}C_{AC},\enskip
\sqrt{\tau^{(i)}_{ABC}}=\alpha^{(i)}\sqrt{\tau_{ABC}}\label{alphabai},\\
p_{(i)}C^{(i)}_{BC}=|k_{(i)}\sqrt{\tau_{ABC}}e^{i(\theta_{(i)}+\tilde{\varphi}_{5})}-C_{BC}b_{(i)}|,\label{cbcchange}\\
p_{(i)}=\lambda^2_{0}a_{(i)}+(1-\lambda^2_{0})b_{(i)}+2\lambda_{0}\lambda_{1}k_{(i)}\cos{(\theta_{(i)}-\varphi)}.\label{p}
\end{eqnarray}
where $C^{(i)}_{AB}$, $C^{(i)}_{AC}$, $C^{(i)}_{BC}$ and $\tau^{(i)}_{ABC}$ are the concurrences and the tangle of the state $\left|\psi^{(i)}\right\rangle$ and 
\begin{eqnarray}
&&M^{\dag}_{(i)}M_{(i)}\equiv\left(
\begin{array}{cc}
a_{(i)} & k_{(i)}e^{-i\theta_{(i)}}  \\
 k_{(i)}e^{i\theta_{(i)}} & b_{(i)} \end{array}
\right),\label{components}\\
&& a_{(i)}b_{(i)}-k^2_{(i)}\ge0,\enskip a_{(i)}\ge0,\enskip b_{(i)}\ge0,\enskip k_{(i)}\ge0,\enskip 0\le\theta_{(i)}\le2\pi,\\
&&\alpha^{(i)}\equiv\frac{\sqrt{a_{(i)}b_{(i)}-k^2_{(i)}}}{p_{(i)}}.\label{alpha}
\end{eqnarray}
\end{lemma}
\textbf{Proof}:
When we perform a transformation which is expressed as 
 \begin{equation}
M=\left(
\begin{array}{cc}
M_{00} & M_{01}  \\
 M_{10} & M_{11} \end{array}
\right),\enskip\enskip\enskip\enskip M_{00}, M_{01}, M_{10}, M_{11}\in \Bbb{C},
\end{equation}
on the qubit $A$ of a pure state \eqref{L2.1}, the state $\left|\psi\right\rangle$ is transformed into
\begin{eqnarray}
M\left|\psi\right\rangle=(\lambda_{0}M_{00}\left|0\right\rangle+\lambda_{0}M_{10}\left|1\right\rangle+\lambda_{1}e^{i\varphi}M_{01}\left|0\right\rangle+\lambda_{1}e^{i\varphi}M_{11}\left|1\right\rangle)\left|00\right\rangle \nonumber \\
+\lambda_{2}(M_{01}\left|0\right\rangle+M_{11}\left|1\right\rangle)\left|01\right\rangle
+\lambda_{3}(M_{01}\left|0\right\rangle+M_{11}\left|1\right\rangle)\left|10\right\rangle+\lambda_{4}(M_{01}\left|0\right\rangle+M_{11}\left|1\right\rangle)\left|11\right\rangle.\label{2012.4.20.1}
\end{eqnarray}
We can transform \eqref{2012.4.20.1} into the form of the generalized Schmidt decomposition \eqref{L2.1} with straightforward algebra;
\begin{eqnarray}%\fl
M\left|\psi\right\rangle=\frac{\lambda_{0}\det{\sqrt{M^{\dag}M}}}{\sqrt{|M_{01}|^2+|M_{11}|^2}}\left|000\right\rangle+\frac{\lambda_{0}(M_{00}M^{*}_{01}+M_{10}M^{*}_{11})+\lambda_{1}e^{i\varphi}(|M_{01}|^2+|M_{11}|^2)}{\sqrt{|M_{01}|^2+|M_{11}|^2}}\left|100\right\rangle \nonumber \\
+\sqrt{|M_{01}|^2+|M_{11}|^2}(\lambda_{2}\left|101\right\rangle+\lambda_{3}\left|110\right\rangle+\lambda_{4}\left|111\right\rangle).\label{11.24.2.22}
\end{eqnarray}
Note that each coefficient of the generalized Schmidt decomposition \eqref{11.24.2.22} of $M\left|\psi\right\rangle$ above is expressed by the components of $M^{\dag}M$ solely:
\begin{eqnarray}%\fl
M^{\dag}M=
\left(
\begin{array}{cc}
|M_{00}|^2+|M_{10}|^2 & M^{*}_{00}M_{01}+M^{*}_{10}M_{11}  \\
 M^{*}_{01}M_{00}+M^{*}_{11}M_{10} & |M_{01}|^2+|M_{11}|^2 \end{array}
\right).
\end{eqnarray}
Then expressing the components of $M^{\dag}_{(i)}M_{(i)}$ as in \eqref{components},
we can also express $M_{(i)}\left|\psi\right\rangle$ and $p_{(i)}$ as
\begin{eqnarray}%\fl
M_{(i)}\left|\psi\right\rangle=\frac{\lambda_{0}\sqrt{a_{(i)}b_{(i)}-k^2_{(i)}}}{\sqrt{b_{(i)}}}\left|000\right\rangle+\frac{\lambda_{0}{k_{(i)}e^{i\theta_{(i)}}}+\lambda_{1}e^{i\varphi}b_{(i)}}{\sqrt{b}_{(i)}}\left|100\right\rangle\nonumber\\+\lambda_{2}\sqrt{b_{(i)}}\left|101\right\rangle+\lambda_{3}\sqrt{b_{(i)}}\left|110\right\rangle+\lambda_{4}\sqrt{b_{(i)}}\left|111\right\rangle. \label{jyunnbi1}
\end{eqnarray}
\begin{eqnarray}
p_{(i)}&=&\left\langle\psi\right|M^{\dag}_{(i)}M_{(i)}\left|\psi\right\rangle\nonumber\\
&=&\lambda^2_{0}a_{(i)}+(1-\lambda^2_{0})b_{(i)}+2\lambda_{0}\lambda_{1}k_{(i)}\cos{(\theta_{(i)}-\varphi)}.\label{L2.24}
\end{eqnarray}
Because of $\left|\psi^{(i)}\right\rangle=M_{(i)}\left|\psi\right\rangle/\sqrt{p_{(i)}}$,
we can express the coefficients of the generalized Schmidt decomposition $\{\lambda^{(i)}|k=0,...,4\}$ as 
\begin{equation}
\lambda^{(i)}_{0}=\frac{\lambda_{0}\sqrt{a_{(i)}b_{(i)}-k_{(i)}^2}}{\sqrt{p_{(i)}}\sqrt{b_{(i)}}}\label{L2.25},
\end{equation}
\begin{equation}
\lambda^{(i)}_{1}e^{i\varphi^{(i)}}=\frac{\lambda_{0}{k_{(i)}e^{i\theta_{(i)}}}+\lambda_{1}e^{i\varphi}b_{(i)}}{\sqrt{p_{(i)}}\sqrt{b_{(i)}}}\label{L2.26},
\end{equation}
\begin{equation}
\lambda^{(i)}_{2}=\frac{\lambda_{2}\sqrt{b_{(i)}}}{\sqrt{p_{(i)}}},\enskip
\lambda^{(i)}_{3}=\frac{\lambda_{3}\sqrt{b_{(i)}}}{\sqrt{p_{(i)}}},\enskip
\lambda^{(i)}_{4}=\frac{\lambda_{4}\sqrt{b_{(i)}}}{\sqrt{p_{(i)}}}.\label{L2.29}
\end{equation}
Because of \eqref{concurrences}, \eqref{tangle} and \eqref{L2.25}--\eqref{L2.29}, the equations \eqref{alphabai} and \eqref{cbcchange} are valid.
$\Box$

\begin{figure}
 \begin{center}
  \includegraphics[width=75mm]{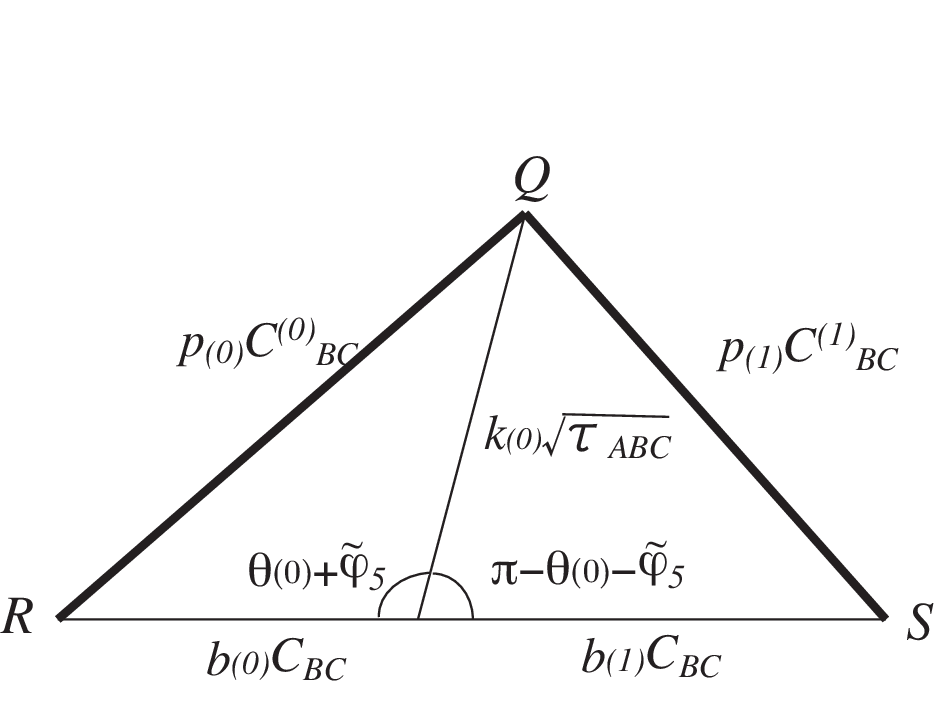}
 \end{center}
 \caption{A geometric interpretation of the change of $C_{BC}$.}
 \label{figlemm2-1}
\end{figure}

\begin{lemma}
Let the notation $\{M_{(i)}|i=1,2\}$ stand for an arbitrary two-choice measurement which is operated on the qubit $A$ of a three-qubit pure state $\left|\psi_{ABC}\right\rangle$. We refer to each result of the measurements $\{M_{(i)}|i=1,2\}$ as $\left|\psi^{(i)}_{ABC}\right\rangle$. 
Let the notations ($C_{AB}$,$C_{AC}$,$C_{BC}$,$\tau_{ABC}$,$J_{5}$,$Q_{\mbox{e}}$) and ($C_{AB}^{(i)}$,$C^{(i)}_{AC}$,$C^{(i)}_{BC}$,$\tau^{(i)}_{ABC}$,$J_{5}^{(i)}$,$Q^{(i)}_{\mbox{e}}$) stand for the sets of the $C$-parameters of the states $\left|\psi_{ABC}\right\rangle$ and $\left|\psi^{(i)}_{ABC}\right\rangle$, respectively. Then, the following inequalities hold:
\begin{equation}
C_{BC}\le \sum_{i=0}^1 p_{(i)}C_{BC}^{(i)} \le \sqrt{C_{BC}^2+\left [1-\left(\sum_{k=0}^1 p_{(i)}\alpha^{(i)}\right)^2\right ]\tau_{ABC}},\label{lemm2target}
\end{equation}
where the probability $p_{(i)}$ and the multiplication factor $\alpha^{(i)}$ are defined in \eqref{p} and \eqref{alpha}, respectively.
\end{lemma}
\textbf{Proof}:
The average $\sum_{i=0}^1 p_{(i)}C_{BC}^{(i)}$ is equal to the length of the heavy line in Fig.~\ref{figlemm2-1}, because we can interpret \eqref{cbcchange} as the cosine theorem and because $b_{(0)}+b_{(1)}=1$ and $\sum_{i}\vec{k}_{(i)}=0$.
The end points of the heavy line have to coincide with the end points of the segment $RS$ because $\sum_{i}\vec{k}_{(i)}=0$. Then, the left inequality $C_{BC}\le \sum_{i=1}^2 p_{i}C_{BC}^{(i)}$ clearly holds, since a polygonal line is longer than a straight line.

To prove the right inequality of this Lemma, it suffices to show the inequality
\begin{eqnarray}
\sqrt{(bC_{BC}+k\cos{\theta}\sqrt{\tau_{ABC}})^2+(k\sin{\theta}\sqrt{\tau_{ABC}})^2}\nonumber\\
+\sqrt{[(1-b)C_{BC}-k\cos{\theta}\sqrt{\tau_{ABC}}]^2+(k\sin{\theta}\sqrt{\tau_{ABC}})^2}\nonumber\\
\le\sqrt{C_{BC}^2+\{1-[\sqrt{ab-k^2}+\sqrt{(1-a)(1-b)-k^2}]^2\}\tau_{ABC}}\label{lemm2targetchange}
\end{eqnarray}
under the conditions $ab-k^2\ge0$, $(1-a)(1-b)-k^2\ge0$, $0\le\theta\le2\pi$, $0\le a\le1$ and $0\le b\le1$, where we used the substitutions:
\begin{equation}
a_{(0)}=a, \enskip a_{(1)}=1-a,\enskip b_{(0)}=b,\enskip b_{(1)}=1-b,\enskip \theta_{(0)}+\tilde{\varphi}_{5}=\pi-\theta.
\end{equation}
The fact that $\theta$ can take any value guarantees the last substitution.

Let us find maximum value of the left-hand side of \eqref{lemm2targetchange} with the values of the measurement parameters $a$, $b$ and $k$ fixed.
We can express the left-hand side of \eqref{lemm2targetchange} as 
\begin{equation}
\sqrt{u^2+w^2+2uw\cos{\theta}}+\sqrt{v^2+w^2-2vw\cos{\theta}},\label{hutosen}
\end{equation}
where $u=bC_{BC}$, $v=(1-b)C_{BC}$ and $w=k\sqrt{\tau_{ABC}}$.
With the values of $u$, $v$ and $w$ fixed, the quantity \eqref{hutosen} is maximized to be $(u+v)\sqrt{1+w^2/(uv)}$ at the point  $\cos{\theta}=w(u-v)/2uv$.
Substituting $u=bC_{BC}$, $v=(1-b)C_{BC}$ and $w=k\sqrt{\tau_{ABC}}$ give that
\begin{equation}
\mbox{the maximum of the left-hand side of \eqref{lemm2targetchange}}=\sqrt{C^2_{BC}+\frac{k^2\tau_{ABC}}{b(1-b)}}.
\end{equation}
Hence, in order to prove Lemma 2, it suffices to show
\begin{equation}
\frac{k^2}{b(1-b)}\le1-[\sqrt{ab-k^2}+\sqrt{(1-a)(1-b)-k^2}]^2.\label{ver4-10}
\end{equation}
After straightforward algebra, \eqref{ver4-10} is reduced to 
\begin{equation}
\left[(a-b)+k^2\frac{(b^2-(1-b)^2)}{b(1-b)}\right]^2\ge0.
\end{equation}
Hence we have proved the right inequality of \eqref{lemm2target} and thereby completed the proof of Lemma 2.$\Box$

Note that Lemma 2 has the following corollary:
\begin{equation}
0\le\sum_{i} p_{(i)}\alpha_{i}\le1.\label{sumalpha}
\end{equation}

Now we have completed the preparation.
Let us start the proof of Theorem 2 in Case $\mathfrak{B}$.
In the present Case, the final state is EP-indefinite, and thus we can neglect the condition 2 of Theorem 2.
Thus, we only have to show that the condition 1 of Theorem 1 is a necessary and sufficient condition of the possibility of a d-LOCC transformation.

Let us prove that a d-LOCC transformation from $\left|\psi\right\rangle$ to $\left|\psi'\right\rangle$ is executable if $\left|\psi\right\rangle$ and $\left|\psi'\right\rangle$ satisfy the condition 1 of Theorem 2.
First we prove that if $\left|\psi\right\rangle$ and $\left|\psi'\right\rangle$ satisfy the condition 1 of Theorem 2, the state $\left|\psi'\right\rangle$ is biseparable or full-separable.
Because $\left|\psi\right\rangle$ is EP-definite and because of \eqref{KABKACKBC},
\begin{equation}
K_{AB}>\tau_{ABC},\enskip K_{AB}>\tau_{ABC},\enskip K_{BC}>\tau_{ABC}.\label{ver4-8}
\end{equation}
Because $\left|\psi'\right\rangle$ is EP-indefinite, at least one of 
\begin{equation}
K'_{AB}=\tau'_{ABC},\enskip K'_{AB}=\tau'_{ABC},\enskip K'_{BC}=\tau'_{ABC}\label{ver4-9}
\end{equation}
holds.
We can assume $K'_{AB}=\tau'_{ABC}$ without losing generality.
Because of \eqref{ver4-8}, $K'_{AB}=\tau'_{ABC}$ and the condition 1 of Theorem 2, at least one of $\zeta$, $\zeta_{A}$, $\zeta_{B}$ is zero.
When only one of $\zeta_{A}$ and $\zeta_{B}$ is zero and $\zeta$ is not zero, the state $\left|\psi'\right\rangle$ is biseparable; only $K'_{BC}$ or $K'_{AC}$ is not zero. 
When both of  $\zeta_{A}$ and $\zeta_{B}$ are zero or $\zeta$ is zero, the state $\left|\psi'\right\rangle$ is full-separable; all of the $K$-parameters of  $\left|\psi'\right\rangle$ are zero. 
Now we have proven that if $\left|\psi\right\rangle$ and $\left|\psi'\right\rangle$ satisfy the condition 1 of Theorem 2, the state $\left|\psi'\right\rangle$ is biseparable or full-separable.

Next, let us prove that if $\left|\psi\right\rangle$ and $\left|\psi'\right\rangle$ satisfy the condition 1 of Theorem 2, there is an executable d-LOCC transformation from $\left|\psi\right\rangle$ to $\left|\psi'\right\rangle$.
Now the state $\left|\psi'\right\rangle$ is biseparable or full-separable.
Without losing generality, we can assume that $C'_{BC}$ is the only nonzero parameter in $(C'_{AB},C'_{AC},C'_{BC},\tau'_{ABC},J'_{5},Q'_{\mbox{e}})$.
Because $\left|\psi\right\rangle$ and $\left|\psi'\right\rangle$ satisfy the condition 1 of Theorem 2, the inequality $C'^2_{BC}\le K_{BC}$ holds.
The set of full-separable states and biseparable states which have the same kind of bipartite entanglement is a totally ordered set \cite{20}.
In other words, when two states $\left|\psi\right\rangle$ and $\left|\psi'\right\rangle$ belong to such a set, there is an executable deterministic LOCC transformation from the EP-definite state $\left|\psi\right\rangle$ to the EP-indefinite state $\left|\psi'\right\rangle$ if and only if the bipartite entanglement of the state $\left|\psi\right\rangle$ is greater than or equal to that of the state $\left|\psi'\right\rangle$.  
Thus, if there is the following measurement $\{M_{(i)}\}$, there is an executable d-LOCC transformation from $\left|\psi\right\rangle$ to $\left|\psi'\right\rangle$;
a measurement whose results can be transformed into a unique state $\left|\psi''\right\rangle$ by local unitary operations without exception, where $\left|\psi''\right\rangle$ is a biseparable state  whose $C^2_{BC}$ is equal to $K_{BC}$ of $\left|\psi\right\rangle$.
The measurement $\{M_{(i)}\}$ is given as follows:
\begin{eqnarray}
M_{(0)}^{\dag}M_{(0)}=\left(
\begin{array}{cc}
a_{(0)} & k_{(0)}e^{-i\theta_{(0)}}  \\
 k_{(0)}e^{i\theta_{(0)}} & b_{(0)} \end{array}
\right)
=\left(
\begin{array}{cc}
  a  & ke^{-i\theta} \\ 
   ke^{i\theta} &  b
\end{array}
\right),\label{surasura5}
\end{eqnarray}
\begin{eqnarray}
M_{(1)}^{\dag}M_{(1)}=\left(
\begin{array}{cc}
a_{(1)} & k_{(1)}e^{-i\theta_{(1)}}  \\
 k_{(1)}e^{i\theta_{(1)}} & b_{(1)} \end{array}
\right)
=\left(
\begin{array}{cc}
  1-a  & -ke^{-i\theta} \\ 
   -ke^{i\theta} &  1-b
\end{array}
\right),
\end{eqnarray}
where the measurement parameters $a$, $b$, $k$ and $\theta$ are defined as follows: 
\begin{equation}
a=\frac{1}{2}-\frac{\lambda_{1}\sin\varphi}{2\sqrt{\lambda^2_{1}\sin^2\varphi+\lambda^2_{0}}},\label{a}
\end{equation}
\begin{equation}
b=\frac{1}{2}+\frac{\lambda_{1}\sin\varphi}{2\sqrt{\lambda^2_{1}\sin^2\varphi+\lambda^2_{0}}},\label{b}
\end{equation}
\begin{equation}
k=\frac{\lambda_{0}}{2\sqrt{\lambda^2_{1}\sin^2\varphi+\lambda^2_{0}}},\label{k}
\end{equation}
\begin{equation}
\theta=\frac{\pi}{2}.\label{theta}
\end{equation}
With substituting \eqref{a}--\eqref{theta} into \eqref{alphabai} and \eqref{cbcchange} and after straightforward algebra, we can confirm that the measurement $\{M_{(i)}\}$ is the measurement that we sought.
Thus, we have proven that the condition 1 of Theorem 2 is a sufficient condition for the existence of an executable d-LOCC transformation.

\begin{figure}
 \begin{center}
  \includegraphics[width=75mm]{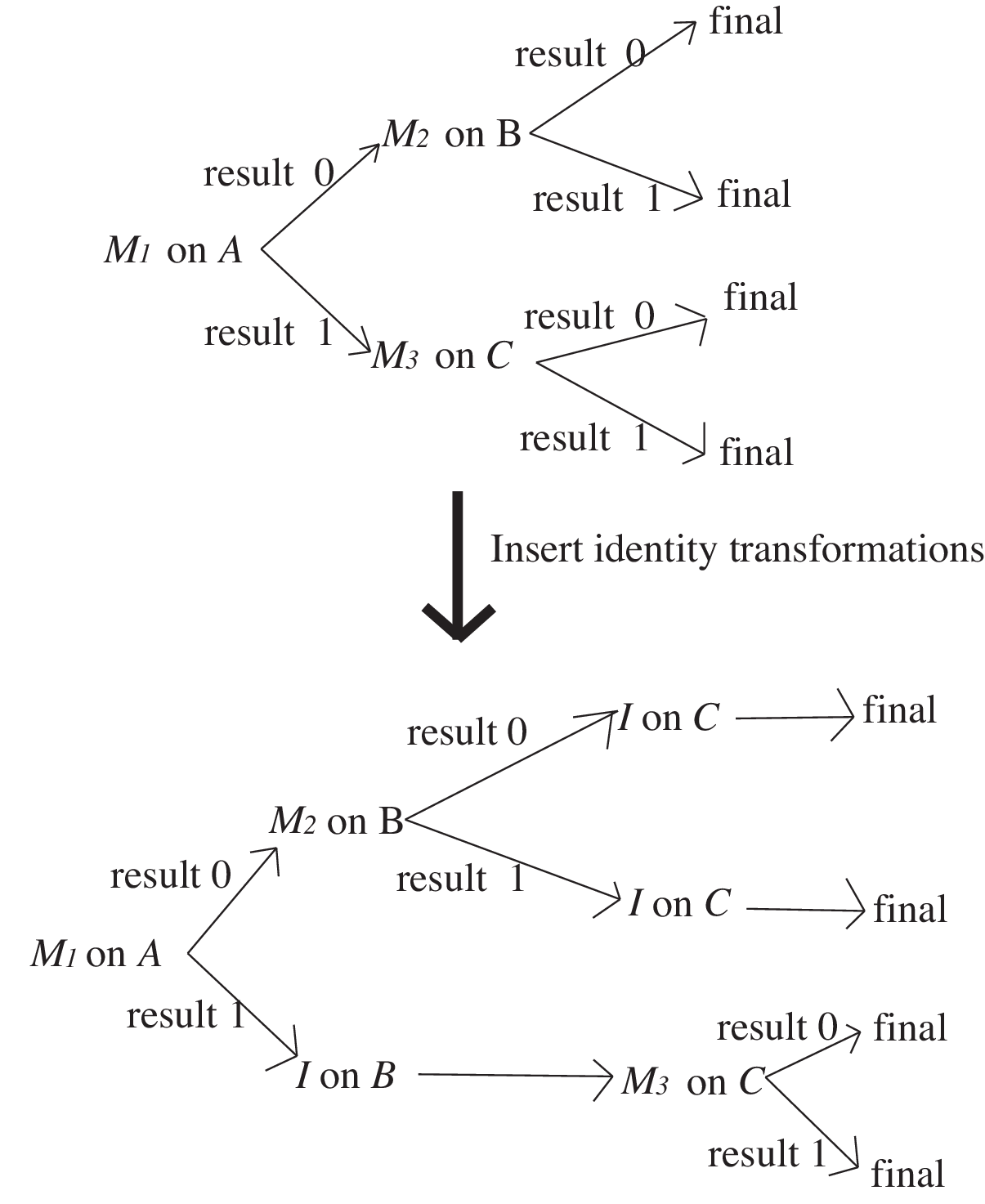}
 \end{center}
 \caption{The method of counting the number $N$. In this figure, $M_{1}$, $M_{2}$ and $M_{3}$ denote measurements and $I$ denotes the identity transformation. The number $N$ is 3 in this example.}
 \label{number}
\end{figure}

Next, let us prove that the condition 1 of Theorem 2 is also a necessary condition.
In other words, we prove that if there is an executable d-LOCC transformation from $\left|\psi\right\rangle$ to $\left|\psi'\right\rangle$, the states $\left|\psi\right\rangle$ and $\left|\psi'\right\rangle$ must satisfy the condition 1 of Theorem 2.
It is possible to substitute two-choice measurements for any measurements of an LOCC transformation on a three-qubit pure state \cite{GHZW}. Hereafter, unless specified otherwise, measurements of LOCC transformations will be two-choice measurements. 
First, we prove that a deterministic LOCC transformation from an EP-definite state to an EP-indefinite state is executable only if the final state is biseparable or full-separable. We prove Lemma 3, which generally holds for stochastic LOCC transformation including d-LOCC transformations. 
\begin{lemma}
Let the notation $T_{SL}$ stand for an LOCC transformation from an arbitrary EP-definite state $\left|\psi\right\rangle$ to arbitrary EP-indefinite states $\{\left|\psi^{(i)}\right\rangle\}$. The subscript SL stands for stochastic LOCC. Then, if this LOCC transformation $T_{SL}$ is executable, there must be full-separable states or biseparable states in the set $\{\left|\psi^{(i)}\right\rangle\}$.
\end{lemma}
\textbf{Proof}:
We prove the present lemma by mathematical induction with respect to $N$, which is the number of times measurements are performed in the LOCC transformation $T_{SL}$. 
Let us define how to count the number of times of the measurement.
Let the notation $\mathbb{T}$ stands for an arbitrary LOCC transformation.
We fix the order of measurements in the LOCC transformation $\mathbb{T}$ cyclically: If the first measurement of the LOCC transformation $\mathbb{T}$ is performed on the qubit $A$, the second one is on the qubit $B$, the third one is on the qubit $C$, the fourth one returns to the qubit $A$, and so on. 
If the first measurement is performed on the qubit $B$, the second one is on the qubit $C$, and so on. 
We can attain such a fixed order by inserting the identity transformation as a measurement. 
The LOCC transformation $\mathbb{T}$ may have branches and the numbers of times the measurements are performed may be different in different branches. We refer to the largest of the numbers as the number $N$. We can make the number of each branch equal to $N$ by inserting the identity transformations.
An example is given in Fig.~\ref{number}.
We use this counting procedure in the proofs of other theorems, too.

Let the notations $(C_{AB},C_{AC},C_{BC},\tau_{ABC},J_{5},Q_{\mbox{e}})$ and $(C^{(i)}_{AB},C^{(i)}_{AC},C^{(i)}_{BC},\tau^{(i)}_{ABC},J^{(i)}_{5},Q^{(i)}_{\mbox{e}})$ stand for the sets of the $C$-parameters of the EP-definite state $\left|\psi\right\rangle$ and the EP-indefinite states $\left|\psi^{(i)}\right\rangle$, respectively.

First, we prove the present lemma for $N=1$. Because of the arbitrariness of the state $\left|\psi\right\rangle$, we can assume that the first measurement $\{M_{(i)}|i=0,1\}$ of the LOCC transformation $T_{SL}$ is performed on the qubit $A$ without loss of generality. 
Thus, the operator $M_{(i)}$ makes $C_{AB}$, $C_{AC}$ and $\sqrt{\tau_{ABC}}$ evenly multiplied by a real number $\alpha^{(i)}$. 
The state $\left|\psi\right\rangle$ is EP definite, and hence $C_{AB}$, $C_{AC}$ and $C_{BC}$ are all positive. 
Because the state $\left|\psi^{(i)}\right\rangle$ is EP indefinite, at least one of $C^{(i)}_{AB}$, $C^{(i)}_{AC}$ and $C^{(i)}_{BC}$ has to be zero for all $i$. 
When $C^{(i)}_{AB}$ or $C^{(i)}_{AC}$ is zero, the multiplication factor $\alpha^{(i)}$ must be zero, and therefore all of $C^{(i)}_{AB}$, $C^{(i)}_{AC}$ and $\tau^{(i)}_{ABC}$ must be zero.  
Then, the parameter $J^{(i)}_{5}$ also must be zero because of $C^{(i)}_{AB}C^{(i)}_{AC}C^{(i)}_{BC}=0$. 
Thus, in the case of $C^{(i)}_{AB}=0$ or $C^{(i)}_{AC}=0$, the EP-indefinite state $\left|\psi^{(i)}\right\rangle$ is a full-separable state with $C^{(i)}_{BC}=0$ or a biseparable state with $C^{(i)}_{BC}\ne0$. Hence, if there were neither a full-separable state nor a biseparable state in the set of EP-indefinite states $\{\left|\psi^{(i)}\right\rangle\}$, the expressions $C^{(i)}_{AC}\ne0$, $C^{(i)}_{AB}\ne0$ and $C^{(i)}_{BC}=0$ would hold for all $i$. 
Because of Lemma 2, however, at least one of $C^{(0)}_{BC}$ and $C^{(1)}_{BC}$ would be greater than or equal to $C_{BC}$, which is positive. 
This is a contradiction, and thus the expression $C^{(i)}_{BC}\ne0$ has to hold for at least one of $i$. We have thereby shown the present lemma for $N=1$.

Now, we prove Lemma 3 for $N=k+1$, assuming that Lemma 3 holds whenever $1\le N\le k$. Let us assume that the number of times of measurements in the LOCC transformation $T_{SL}$ from the EP-definite state $\left|\psi\right\rangle$ to the EP-indefinite states $\{\left|\psi^{(i)}\right\rangle\}$ is $k+1$. 
Because of the assumption for $1\le N\le k$, the situation before the last measurement has to be either of the following two situations: 
\begin{description}
\item[(i)]All states are already EP indefinite, and there are full-separable states or biseparable states among them.  
\item[(ii)]Some states are EP definite.
\end{description}
In the case of (i), there are full-separable states or biseparable states in the final EP-indefinite states $\{\left|\psi^{(i)}\right\rangle\}$ because an arbitrary full-separable state or an arbitrary biseparable state can be transformed only into full-separable states or biseparable states by a measurement.$\\$
In the case of (ii), if there were neither a full-separable state nor a biseparable state in the EP-indefinite states $\{\left|\psi^{(i)}\right\rangle\}$, there would have to be a measurement which could transform an EP-definite state to EP-indefinite states which are neither full-separable states nor biseparable states. Because of the theorem for $N=1$, this is impossible. 

Therefore, there must be either full-separable states or biseparable states in the EP-indefinite states $\{\left|\psi^{(i)}\right\rangle\}$ in the case (ii) as well as in the case (i). This completes the proof of Lemma 3. $\Box$
$\\$

Because of Lemma 3, if a d-LOCC from an EP-definite state $\left|\psi\right\rangle$ to an EP-indefinite state $\left|\psi'\right\rangle$ is executable, then $\left|\psi'\right\rangle$ is biseparable or full-separable.
Without losing generality, we can assume that the only nonzero parameter in the $K$-parameters of $\left|\psi'\right\rangle$ is $K'_{BC}$.
To show the $K$-parameters of　$\left|\psi\right\rangle$ and $\left|\psi'\right\rangle$ satisfy the condition 1 of Theorem 2, it is sufficient to prove the inequality $K'_{BC}\le K_{BC}$.
To prove $K'_{BC}\le K_{BC}$, we prove the following Lemma 4.
\begin{lemma}
Let the notation $\{M_{(i)}|i=0,1\}$ stand for an arbitrary two-choice measurement which is operated on a qubit of a three-qubit pure state $\left|\psi_{ABC}\right\rangle$.
Note that we can operate  $\{M_{(i)}|i=0,1\}$ on any one of the qubits $A$, $B$ and $C$ of the state $\left|\psi_{ABC}\right\rangle$.
We refer to each result of $\{M_{(i)}|i=0,1\}$ as $\left|\psi^{(i)}_{ABC}\right\rangle$. 
Let the notations ($K_{AB}$,$K_{AC}$,$K_{BC}$,$\tau_{ABC}$,$J_{5}$,$Q_{\mbox{e}}$) and ($K_{AB}^{(i)}$,$K^{(i)}_{AC}$,$K^{(i)}_{BC}$,$\tau^{(i)}_{ABC}$,$J_{5}^{(i)}$,$Q^{(i)}_{\mbox{e}}$) stand for the sets of the $K$-parameters of the states $\left|\psi_{ABC}\right\rangle$ and $\left|\psi^{(i)}_{ABC}\right\rangle$, respectively. Then, the following inequality holds:
\begin{equation}
\sum_{i=0}^1 p_{(i)}\sqrt{K^{(i)}_{BC}}\le\sqrt{K_{BC}}.\label{theo4hutou}
\end{equation}
\end{lemma}

\textbf{Proof}: First, we prove \eqref{theo4hutou} in the case where the measurement $\{M_{(i)}|i=0,1\}$ is performed on the qubit $B$ or $C$. In this case $C^{(i)}_{BC}=\alpha^{(i)}C_{BC}$ and $\sqrt{\tau^{(i)}_{ABC}}=\alpha^{(i)}\sqrt{\tau_{ABC}}$, where $\alpha^{(i)}$ is the multiplication factor of the measurement $\{M_{(i)}|i=0,1\}$. Thus, because of $\eqref{sumalpha}$, we can obtain $\eqref{theo4hutou}$ as follows:
\begin{equation}%\fl
\sum_{i=0}^1 p_{(i)}\sqrt{K^{(i)}_{BC}}=\sum_{i=0}^1 p_{(i)}\sqrt{(C^{(i)}_{BC})^2+\tau^{(i)}_{ABC}}=\sum_{i=0}^1 p_{(i)}\alpha^{(i)}\sqrt{C^2_{BC}+\tau_{ABC}}\le\sqrt{K_{BC}}.
\end{equation}

Now, it suffices to prove $\eqref{theo4hutou}$ in the case where the first measurement is performed on the qubit $A$.  
Let the notation $f$ stand for the left-hand side of $\eqref{theo4hutou}$.
Because of Lemma 1 we obtain
\begin{eqnarray}%\fl
f&=&\sqrt{b^2C^2_{BC}+2bk\cos\theta C_{BC}\sqrt{\tau_{ABC}}+ab\tau_{ABC}}\\%\fl
&&+\sqrt{(1-b)^2C^2_{BC}-2(1-b)k\cos\theta C_{BC}\sqrt{\tau_{ABC}}+(1-a)(1-b)\tau_{ABC}},\nonumber
\end{eqnarray}
where we substitute $\theta$ for the phase $\pi-\theta-\tilde{\varphi}_{5}$, because the range of the phase $\theta$ is from 0 to $2\pi$.
When $2k\cos\theta C_{BC}=(b-a)\sqrt{\tau_{ABC}}$, the function $f$ is maximized to $\sqrt{K_{BC}}$.
Thus, $\eqref{theo4hutou}$ holds.
$\Box$

The inequality \eqref{theo4hutou} includes $K'_{BC}\le K_{BC}$, if the number $N$ of times measurements of a d-LOCC transformation from $\left|\psi\right\rangle$ to $\left|\psi'\right\rangle$ is equal to 1.
Let us assume that $K'_{BC}\le K_{BC}$ also holds whenever $1\le N\le k$.
Then, when $N=k+1$, the inequality $K'_{BC}\le K_{BC}$ also holds;
$(Proof$: Let the notion $\{\left|\psi'^{(i)}\right\rangle\}$ stand for results of the first measurement the d-LOCC transformation.
We refer to the parameter $K_{BC}$ of $\left|\psi'^{(i)}\right\rangle$ as $K'^{(i)}_{BC}$.
Note that the LOCC transformations from $\left|\psi'^{(i)}\right\rangle$ to $\left|\psi'\right\rangle$
are d-LOCC transformations whose the number of times measurements are less than or equal to $k$.
Thus, the inequalities $K'_{BC}\le K'^{(i)}_{BC}$ hold for all of $K'^{(i)}_{BC}$, and thus the inequality $K'_{BC}\le K_{BC}$ also holds.
$\Box)$
Hence, now we have completed the proof of Theorem 2 in Case $\mathfrak{B}$.

\subsection{Case $\mathfrak{C}$}
In this subsection, we prove Main Theorems in Case $\mathfrak{C}$, where the initial state is EP-indefinite and GHZ-type.

First, we consider the case in which the final state is EP-definite.
In this case, the final state is truly multipartite.
We cannot achieve a W-type state from a GHZ-type state with an LOCC transformation \cite{GHZW}.
Thus, the final state must be a GHZ-type state too.
We have already proven that if the initial and final states satisfy the two conditions of Theorem 2, the final state must be GHZ-type. 
Thus, we only have to consider the case in which both initial and final states are GHZ-type. 
A d-LOCC transformation from a GHZ-type state whose coefficient set $\{c_{i}\}$ has a zero to a GHZ-type state whose coefficient set $\{c_{i}\}$ has no zeros is executable if and only if 
\begin{eqnarray}
c'_{i}\ge c_{i}\enskip(k=A,B,C),\label{revise17}\\
|z|=1,\label{revise18}\\
\mbox{$z'$ is purely imaginary},\label{revise19}
\end{eqnarray}
where $\{c_{i}\}$ and $z$ are defined in \eqref{cdef} \cite{25}.
Because of \eqref{revise4}, a GHZ-type state is EP-definite if and only if its coefficient set $\{c_{i}\}$ has no zeros.
Thus, in the present case where the final state is EP-definite, we only have to prove that \eqref{revise17}--\eqref{revise19} are equivalent to the two conditions of Theorem 2. 

Let us prove the equivalence between \eqref{revise17}--\eqref{revise19} are the two conditions of Theorem 2. 
Because of \eqref{revise11} and \eqref{revise4}, the expressions \eqref{revise17}--\eqref{revise19} are equivalent to
\begin{eqnarray}
\frac{\tau'_{ABC}}{\tau_{ABC}}\le\frac{K'_{AB}}{K_{AB}},\enskip\frac{\tau'_{ABC}}{\tau_{ABC}}&\le&\frac{K'_{AC}}{K_{AC}},\enskip\frac{\tau'_{ABC}}{\tau_{ABC}}\le\frac{K'_{BC}}{K_{BC}},\label{revise20}\\
\frac{\sqrt{K_{AB}K_{AC}}}{2\lambda^2_{0}\sqrt{K_{BC}}}&=&\frac{\sqrt{K_{\mbox{ap}}}}{K_{5}\pm\sqrt{\Delta_{J}}}=1,\label{revise21}\\
\tilde{\varphi}_{5}&=&\pm\frac{\pi}{2},\label{revise22}
\end{eqnarray}
where the double sign $\pm$ in \eqref{revise21} is the one in \eqref{ambi1}; note that the possible two sets of $\{\lambda_{i},\varphi|i=0,...,1\}$ are $\{\lambda^{+}_{i},\varphi^{+}|i=0,...,1\}$ 
and $\{\lambda^{-}_{i},\tilde{\varphi}^{-}|i=0,...,1\}$ or
 $\{\lambda^{-}_{i},\varphi^{-}|i=0,...,1\}$ and 
 $\{\lambda^{+}_{i},\tilde{\varphi}^{+}|i=0,...,1\}$.
Because \eqref{revise21} is valid in either case of the multiple signs, the parameter $\Delta_{J}$ is zero.
Because of $\Delta_{J}=K^2_{5}-K_{\mbox{ap}}$, when $\Delta_{J}=0$ holds, \eqref{revise21} also holds.
Thus, \eqref{revise21} is equivalent to $\Delta_{J}=0$.
Because of $\sqrt{J_{\mbox{ap}}}\cos\tilde{\varphi}_{5}=J_{5}$, \eqref{revise22} is equivalent to $J_{5}=J_{\mbox{ap}}=0$.
Thus, \eqref{revise21} and \eqref{revise22} are equivalent to \eqref{2012.1}.
We have already proven that \eqref{2012.1} is equivalent to $\zeta_{\mbox{lower}}\le\zeta\le1$ and \eqref{revise15} in the section 4.1.
Thus, we only have to prove that \eqref{revise20} is equivalent to the existence of $\zeta$, $0\le\zeta_{A}\le1$, $0\le\zeta_{B}\le1$ and $0\le\zeta_{C}\le1$, which satisfy \eqref{lastlast}.
We can define such $\zeta$--$\zeta_{C}$ as \eqref{revise23} and \eqref{revise24}.
Thus, in the case where the final state is EP-definite, the conditions of Theorem 2 is a necessary and sufficient condition of d-LOCC.

Second, we consider the case where the final state is EP-indefinite and GHZ-type.
Because a GHZ-type state is EP-definite if and only if its coefficient set $\{c_{i}\}$ has no zeros, in this case,
a d-LOCC transformation is executable if and only if 
\begin{eqnarray}
c'_{i}\ge c_{i}\enskip(k=A,B,C),\label{revise25}\\
|z'|\ge|z|,\label{revise26}
\end{eqnarray}
where $\{c_{i}\}$ and $z$ are defined in \eqref{cdef}, and where we choose $z$ and $z'$ such that $|z|\ge1$ and $|z'|\ge1$ \cite{25}.

Let us prove that \eqref{revise25} and \eqref{revise26} are equivalent to the condition 1 of Theroem 2.
Because of \eqref{ambi1}, \eqref{revise11}, $|z'|\ge1$ and $|z|\ge1$, 
\begin{equation}
|z|=\frac{\sqrt{K_{\mbox{ap}}}}{K_{5}-\sqrt{\Delta_{J}}},\enskip|z'|=\frac{\sqrt{K'_{\mbox{ap}}}}{K'_{5}-\sqrt{\Delta'_{J}}}.\label{revise27}
\end{equation}
Because of \eqref{revise4}, \eqref{revise25} is equivalent \eqref{revise20}.
Because both of the initial and final states are EP-indefinite, $J_{5}=J'_{5}=0$ is valid.
Thus, we can define $\zeta_{\mbox{lower}}\le\zeta\le1$, $0\le\zeta_{A}\le1$, $0\le\zeta_{B}\le1$ and $0\le\zeta_{C}\le1$, which satisfy \eqref{lastlast} as \eqref{revise23} and \eqref{revise24}.
Thus, in the case where the final state is EP-indefinite and GHZ-type, the conditions of Theorem 2 is a necessary and sufficient condition of d-LOCC.

Third, we consider the case where the final state is EP-indefinite but not GHZ-type.
In this case, the final state is biseparable or full-separable.
Because the final state is not EP-definite, the condition 2 of Theorem 2 is left out; we only have to consider the condition 1.
When the final state is full-separable, a d-LOCC transformation to the final state is clearly executable, and the initial and final states clearly satisfy the condition 1 of Theorem 2: $\zeta_{A}=\zeta_{B}=\zeta_{C}=0$ satisfy \eqref{lastlast}. 
Thus, we only have to consider the case the final state is biseparable.

Let us prove the condition 1 of Theorem 2 is a necessary condition of the possibility of the d-LOCC transformation.
Without loss of generality, we can assume that the only nonzero $K$-parameter of the final state is $K'_{BC}$.
Because of Lemma 4, we can prove the inequality $K'_{BC}\le K_{BC}$, in the same manner as $K'_{BC}\le K_{BC}$ is shown in Case $\mathfrak{B}$.
Thus, if a d-LOCC transformation is executable, the initial and final states satisfy the condition 1 of Theorem 2.
In other words, the condition 1 of Theorem 2 is a necessary condition of the possibility of the d-LOCC transformation.

Finally, let us prove the condition 1 of Theorem 2 is a sufficient condition of the possibility of the d-LOCC transformation.
Without loss of generality, we can assume that the only nonzero $K$-parameter of the final state is $K'_{BC}$.
The upper limit of $K'_{BC}$ is $K_{BC}$.
We can realize this upper limit by the measurement $\{M_{(i)}\}$ which is defined as \eqref{surasura5}--\eqref{theta}.
According to Ref. \cite{20}, the set of full-separable states and biseparable states which have the same kind of bipartite entanglement is a totally ordered set.
Thus, in this case, a d-LOCC transformation from $\left|\psi\right\rangle$ to $\left|\psi'\right\rangle$ is executable if and only if $K'_{BC}\le K_{BC}$.
Thus, in the case where the final state is EP-indefinite and not GHZ-type, the conditions of Theorem 2 is a necessary and sufficient condition of the possibility of d-LOCC transformation.
$\Box$

\subsection{Case $\mathfrak{D}$}
In this subsection, we prove Theorem 1 in Case $\mathfrak{D}$, where the tangle of the initial state is zero.

First, we simplify $\eqref{lastlast}$.
Let us show that we can leave $\tau_{ABC}$, $J_{5}$ and $Q_{\mbox{e}}$ out of the discussion hereafter.
First, $\tau'_{ABC}=0$ follows from $\tau_{ABC}=0$, because an arbitrary measurement makes the tangle $\tau_{ABC}$ only multiplied by a constant. 
Next, because of $\eqref{concurrences}$--$\eqref{J5}$ and the equation $\tau_{ABC}=0$, the equation $J_{5}=C_{AB}C_{AC}C_{BC}$ holds.
$(Proof$: 
Because of \eqref{tangle} and $\tau_{ABC}=0$, at least one of $\lambda_{0}$ and $\lambda_{4}$ is zero.
When $\lambda_{0}$ is zero, the equations $J_{5}=C_{AB}C_{AC}C_{BC}=0$ hold.
Thus $\lambda_{4}$ is zero, and then $J_{5}=C_{AB}C_{AC}C_{BC}=4\lambda^2_{0}\lambda^2_{2}\lambda^2_{3}$ follows \eqref{concurrences} and \eqref{J5}.$\Box)$ 
Thus, in order to examine the change of  $J_{5}$, it suffices to examine the change of the concurrences $C_{AB}$, $C_{AC}$ and $C_{BC}$. 

Next, because of $\tau_{ABC}=0\Leftrightarrow\lambda_{0}=0\lor\lambda_{4}=0$ and because if there is a zero in $\{\lambda_{k}|k=0,...,4\}$ then $\sin\varphi=0$, the equation $Q_{\mbox{e}}=0$ follows \eqref{Qe}.
In the same manner, the entanglement charge $Q'_{\mbox{e}}$ is also zero because of $\tau'_{ABC}=0$.
Thus, $Q_{\mbox{e}}=Q'_{\mbox{e}}=0$ holds.
Condition 2 of Theorem 2 satisfies this equation.
Let us show this. The state $\left|\psi\right\rangle$ is $\tilde{\zeta}$-indefinite, because $\tau_{ABC}=0$:
\begin{eqnarray}
\Delta_{J}=K^2_{5}-K_{\mbox{ap}}=J^2_{5}-J_{\mbox{ap}}=C^2_{AB}C^2_{AC}C^2_{BC}\sin^2\varphi_{5}\nonumber\\
=4C^2_{AB}C^2_{AC}\lambda^2_{1}\lambda^2_{4}\sin^2\varphi=0,
\end{eqnarray}
where we use $\sin\varphi=0$.
Note that $\sin\varphi=0$ holds when there is a zero in $\{\lambda_{i}|i=0,...,4\}$ and that $\tau_{ABC}=\lambda_{0}\lambda_{4}=0$.
Because the state $\left|\psi\right\rangle$ is $\tilde{\zeta}$-indefinite, Condition 2 is reduced to 
 $|Q'_{\mbox{e}}|=\mbox{sgn}[(1-\zeta)(\zeta-\zeta_{\mbox{lower}})]$.
Incidentally, because of $\tau_{ABC}=0$, $\zeta_{\mbox{lower}}=1$ holds.
Thus, $\zeta=1$ follows $\zeta_{\mbox{lower}}\le\zeta\le1$, and thus Condition 2 satisfies the equation $Q'_{\mbox{e}}=0$.
In order to prove the present theorem, it suffices to show that a necessary and sufficient condition is that there are real numbers $\alpha_{A}$, $\alpha_{B}$ and $\alpha_{C}$ which are from zero to one and which satisfy the following equation:
\begin{equation}
\left(
\begin{array}{c}
C'^2_{AB} \\
C'^2_{AC} \\
C'^2_{BC} \\
\end{array}
\right)=
\left(
\begin{array}{ccc}
\alpha_{A}^2\alpha_{B}^2 &   &      \\
  & \alpha_{A}^2\alpha_{C}^2 &    \\
  &   & \alpha_{B}^2\alpha_{C}^2  \\
\end{array}
\right)\left(
\begin{array}{c}
C_{AB}^2 \\
C_{AC}^2 \\
C_{BC}^2 \\
\end{array}
\right).\label{180.10.26}
\end{equation}
Note that \eqref{180.10.26}, $J_{5}=C_{AB}C_{AC}C_{BC}$ and $J'_{5}=C'_{AB}C'_{AC}C'_{BC}$ give $J'_{5}=\alpha^2_{A}\alpha^2_{B}\alpha^2_{C}J_{5}$, and that $\tau_{ABC}=0\Rightarrow (K_{AB}=C^2_{AB})\land(K_{AC}=C^2_{AC})\land(K_{BC}=C^2_{BC})$.

We can classify the sets of the initial and final states as follows:
\begin{description}
\item[Case $\mathfrak{D}$-1:]At least one of the concurrences $C_{AB}$, $C_{BC}$ and $C_{AC}$ is zero. 
\item[Case $\mathfrak{D}$-2:]None of the concurrences $C_{AB}$, $C_{BC}$ and $C_{AC}$ is zero, and at least one of the concurrences $C'_{AB}$, $C'_{BC}$ and $C'_{AC}$ is zero.
\item[Case $\mathfrak{D}$-3:]All of the concurrences $C_{AB}$, $C_{BC}$, $C_{AC}$, $C'_{AB}$, $C'_{BC}$ and $C'_{AC}$ are nonzero.
\end{description}
Note that we already have $\tau_{ABC}=\tau'_{ABC}$ in the present Case $\mathfrak{D}$. 

In Case $\mathfrak{D}$-1, we first note that only the biseparable states are allowed as the initial states in this case.
The set of full-separable states and biseparable states which have the same kinds of bipartite entanglements is a totally ordered set \cite{20}.
We cannot transform a full-separable state or a biseparable state into other type states with LOCC transformations\cite{GHZW}, and thus we can derive the necessary and sufficient condition, which reduces to the following: there is an executable deterministic LOCC transformation from $\left|\psi\right\rangle$ to $\left|\psi'\right\rangle$ if and only if $C_{AC}\ge C'_{AC}$.
This condition is equivalent to \eqref{180.10.26} in Case $\mathfrak{D}$-1.

In Case $\mathfrak{D}$-2, where none of $C_{AB}$, $C_{BC}$ and $C_{AC}$ is zero and at least one of  $C'_{AB}$, $C'_{BC}$ and $C'_{AC}$ is zero, the initial state is EP definite while the final state is EP indefinite.
Thus, Case $\mathfrak{B}$ includes Case $\mathfrak{D}$-2, and thus the existence of $\alpha_{A}$, $\alpha_{B}$ and $\alpha_{C}$ which satisfy \eqref{180.10.26} is the necessary and sufficient condition.

In Case $\mathfrak{D}$-3, where all of the concurrences $C_{AB}$, $C_{BC}$, $C_{AC}$, $C'_{AB}$, $C'_{BC}$ and $C'_{AC}$ are nonzero, the initial and final states are W-type states.
In the present case, we can use the result of Ref. \cite{30}.
A d-LOCC transformation from a W-type state $\left|\psi\right\rangle$ to another W-type state $\left|\psi'\right\rangle$ is possible if and only if  
\begin{equation}
x_{i}\ge x'_{i}\enskip(i=1,2,3),\label{revise3}
\end{equation}
where the sets of positive real numbers $\{x_{i}\}$ and $\{x'_{i}\}$ are defined by the decompositions of $\left|\psi\right\rangle$ and $\left|\psi'\right\rangle$ \cite{30}:
\begin{eqnarray}
\left|\psi\right\rangle=x_{0}\left|000\right\rangle+x_{1}\left|100\right\rangle+x_{2}\left|010\right\rangle+x_{3}\left|001\right\rangle\label{revise1},\\
\left|\psi'\right\rangle=x'_{0}\left|000\right\rangle+x'_{1}\left|100\right\rangle+x'_{2}\left|010\right\rangle+x'_{3}\left|001\right\rangle.
\end{eqnarray}
Note that we can reduce \eqref{revise1} into a generalized Schmidt decomposition
\begin{equation}
\left|\psi\right\rangle=x_{1}\left|000\right\rangle+x_{0}\left|100\right\rangle+x_{3}\left|101\right\rangle+x_{2}\left|110\right\rangle
\end{equation}
with transformation $\left|0_{A}\right\rangle\leftrightarrow\left|1_{A}\right\rangle$.  
We thereby obtain
\begin{eqnarray}
2x_{1}x_{2}=C_{AB},\enskip 2x_{1}x_{3}=C_{AC},\enskip 2x_{2}x_{3}=C_{BC},\\
x_{1}=\sqrt{\frac{C_{AB}C_{AC}}{2C_{BC}}},\enskip x_{2}=\sqrt{\frac{C_{AB}C_{BC}}{2C_{AC}}},\enskip x_{1}=\sqrt{\frac{C_{AC}C_{BC}}{2C_{AB}}}.    
\end{eqnarray}
Thus, the existence of $\alpha_{A}$, $\alpha_{B}$ and $\alpha_{C}$ which are from zero to one which satisfy \eqref{180.10.26} is equivalent to the existence $\alpha_{A}$, $\alpha_{B}$ and $\alpha_{C}$ which are from zero to one which satisfy
\begin{equation}
\alpha_{A}=x'_{1}/x_{1},\enskip \alpha_{B}=x'_{2}/x_{2},\enskip \alpha_{C}=x'_{3}/x_{3}.\label{revise2}
\end{equation}
Note that when $\alpha_{A}$, $\alpha_{B}$ and $\alpha_{C}$ are from 0 to 1, \eqref{revise2} is equivalent to \eqref{revise3}.
Thus, \eqref{180.10.26} is a necessary and sufficient condition of d-LOCC in Case $\mathfrak{D}$-3.
$\Box$
  
We thereby have completed the proof of Theorem 2 in all cases.

\section{Conclusion}
In the present paper, we have given four important results.
First, we have introduced the entanglement charge $Q_{\mbox{e}}$.
This new entanglement parameter $Q_{\mbox{e}}$ has features which the electric charge has.
The set of the six parameters $(C_{AB}, C_{AC}, C_{BC}, \tau_{ABC}, J_{5}, Q_{\mbox{e}})$ is a complete set for the LU-equivalence;  arbitrary three qubit pure states are LU-equivalent if and only if their entanglement parameters $(C_{AB}, C_{AC}, C_{BC}, \tau_{ABC}, J_{5}, Q_{\mbox{e}})$ are equal to each other.
This result means that the nonlocal features of three-qubit pure states can be expressed completely in terms of the magnitudes, phase and charge of the entanglement.
The entanglement charge $Q_{\mbox{e}}$ satisfies a conservation law partially.
Deterministic LOCC transformations between $\tilde{\zeta}$-definite states conserve the entanglement charge $Q_{\mbox{e}}$.
When we transform a $\tilde{\zeta}$-indefinite state into a $\tilde{\zeta}$-definite state, we can choose the value of the entanglement charge. Once the value is determined, we cannot change it anymore (Fig.~\ref{denkakouzou}).
In this sense, we can regard $\tilde{\zeta}$-indefinite states as charge-definite states, and a deterministic LOCC transformation between charge-definite states preserves the entanglement charge.

Second, we have given a necessary and sufficient condition of the possibility of deterministic LOCC transformations of three-qubit pure states.
The necessary and sufficient condition is given as a condition between the vectors $(K_{AB}, K_{AC}, K_{BC}, \tau_{ABC}, J_{5}, Q_{\mbox{e}})$ of the initial and final states of deterministic LOCC transformation.
In other words, we have revealed that three-qubit pure states are a partially ordered set parametrized by the six entanglement parameters. 
Note that other multipartite systems may have similar structures;
it is plausible that the nonlocal features of $N$-qubit pure states can be expressed completely in terms of the magnitudes, phases and charges of the entanglement.
Then the approach of the present paper may be applicable to such systems. 

Third, we have clarified the the rule how a deterministic measurement changes the entanglement. We can express this change as in Fig.~\ref{ETransfer}.
The rule indicates the transfer of the entanglement.
After performing a deterministic measurement on the qubit $A$, the four entanglement parameters, $C^2_{AB}$, $C^2_{AC}$, $\tau^2_{ABC}$ and $J_{5}$ are multiplied by $\alpha^2_{A}$.  
The quantity $\beta_{A}(1-\alpha^2_{A})\tau_{ABC}$, which is a part of the entanglement lost from $\tau_{ABC}$, is added to $C^2_{BC}$. 
The quantity $(1-\beta_{A})(1-\alpha^2_{A})\tau_{ABC}$, which is the rest of the entanglement lost from $\tau_{ABC}$ disappear.
We call this phenomenon the entanglement transfer.

Fourth, we have given the minimum times of measurements to reproduce an arbitrary executable deterministic LOCC transformation. We can realize the minimum times by performing DMTs. We can also determine the order of measurements; we can determine which qubit is measured first, second and third.

Is there entanglement transfer for a stochastic LOCC transformation? 
For this question, the present paper has given a partial answer. 
Let us see the inequalities given in Lemma 2:
\begin{equation}
C_{BC}\le \sum_{i=1}^2 p_{(i)}C_{BC}^{(i)} \le \sqrt{C_{BC}^2+\left[1-\left(\sum_{k=1}^2 p_{i}\alpha^{(i)}\right)^2\right]\tau_{ABC}}\label{11.11.9.1}.
\end{equation}
The left inequality means that the bipartite entanglement $C_{BC}$ between the qubits $B$ and $C$ increases when the qubit $A$ is measured. 
The right inequality is equivalent to the following inequality:
\begin{equation}
\left(\sum_{i=1}^2 p_{(i)}C_{BC}^{(i)}\right)^2+\left(\sum_{i=1}^2 p_{(i)}\sqrt{\tau_{ABC}^{(i)}}\right)^2 \le C_{BC}^2+\tau_{ABC},\label{12.6.10.2}
\end{equation}
because $\sqrt{\tau^{(i)}_{ABC}}=\alpha_{(i)}\sqrt{\tau_{ABC}}$.
We can interpret the left-hand side of $\eqref{12.6.10.2}$ as the sum of the bipartite entanglement $C_{BC}$ between the qubits $B$ and $C$ and the tripartite entanglement $\tau_{ABC}$ among the qubits $A$, $B$ and $C$ \textit{after} a measurement.
On the other hand, the right-hand side is the sum \textit{before} a measurement. 
Thus, $\eqref{12.6.10.2}$ means that a measurement decreases the sum.
Note that the bipartite entanglement $C_{BC}$ of the qubits $B$ and $C$ increases, whereas the tripartite entanglement $\tau_{ABC}$ among the qubits $A$, $B$ and $C$ decreases.
To summarize the above, a kind of dissipative entanglement transfer also occurs for a two-choice measurement which are not a deterministic measurement.
It is expected that the transfer occurs for an $n$-choice measurement too.
Indeed, the left inequality of $\eqref{11.11.9.1}$ also holds for an $n$-choice measurement.
However, the right inequality of $\eqref{11.11.9.1}$ for an $n$-choice measurement has not been proven yet.

In  the present paper, we have exhaustively analyzed deterministic LOCC transformations of three-qubit pure states.
This is the first step of the extension of Nielsen's work \cite{20} to multipartite entanglements.

\section*{Acknowledgements}
 
The present work is supported by CREST of Japan Science and Technology Agency.
The author is indebted to Dr. Julio de Vicente for his valuable comments. 
The author also thanks Prof. Naomichi Hatano and Prof. Satoshi Ishizaka for useful discussions.
In particular, Prof. Naomichi Hatano patiently checked this very long thesis and give the author many pieces of important advice.

\appendix
\section{The proof that $Q_{\mbox{e}}$ is a tripartite parameter}

In the present section, we show that $Q_{\mbox{e}}$ defined in \eqref{Qe} is a tripartite parameter; in other words, we show that $Q_{\mbox{e}}$ is invariant with respect to permutations of the qubits $A$, $B$ and $C$.
First, we perform the proof in the case of $Q_{\mbox{e}}=0$.
Because of \eqref{ambi1} and \eqref{Qe}, the equation $Q_{\mbox{e}}=0$ holds if and only if $\Delta_{J}=0\lor \sin\varphi=0$.
Because of \eqref{concurrences} and \eqref{J5}, the expression $\sin\varphi=0$ is equivalent to $|J_{5}|=C_{AB}C_{AC}C_{BC}$.
Thus, $Q_{\mbox{e}}=0$ is equivalent to $\Delta_{J}=0\lor |J_{5}|=C_{AB}C_{AC}C_{BC}$.
Therefore, if we can show that the expression $\Delta_{J}=0\lor |J_{5}|=C_{AB}C_{AC}C_{BC}$ is invariant with respect to permutations of $A$, $B$ and $C$, we can also show that $Q_{\mbox{e}}$ is invariant with respect to the permutations.
The parameters $J_{5}$ and $C_{AB}C_{AC}C_{BC}$ are invariant with respect to the permutations of $A$, $B$ and $C$ \cite{18}.
This fact and \eqref{Delta_{J}} give that $\Delta_{J}$ is also invariant with respect to the permutations of $A$, $B$ and $C$. 
Hence, the quantities $\Delta_{J}$, $J_{5}$ and $C_{AB}C_{AC}C_{BC}$ are invariant with respect to the permutations of $A$, $B$ and $C$, and thus if $Q_{\mbox{e}}=0$, then $Q_{\mbox{e}}$ is invariant with respect to permutations of $A$, $B$ and $C$.
Namely, if $Q_{\mbox{e}}=0$, then $Q_{\mbox{e}}$ is a tripartite parameter.

Second, we perform the proof in the case of  $Q_{\mbox{e}}=\pm1$.
In order to show this, we only have to show that $Q_{\mbox{e}}$ is invariant with respect to the permutation of $A$ and $B$, because if we can prove the invariance with respect to the permutation of $A$ and $B$ we can also prove the invariance with respect to the permutation of $A$ and $C$ or $B$ and $C$ in the same manner.

Let us derive the generalized Schmidt decomposition whose order of the qubits is $BAC$ and see the expression of $Q_{\mbox{e}}$ in the new decomposition, which we refer to as $Q^{B}_{\mbox{e}}$. 
We can assume that \eqref{L2.1} is a positive decomposition and let us permute $A$ and $B$ of \eqref{L2.1}:
\begin{equation}%\fl
\left|\psi\right\rangle=\lambda_{0}\left|0_B0_A0_C\right\rangle+\lambda_{1}e^{i\varphi}\left|0_B1_A0_C\right\rangle+\lambda_{2}\left|0_B1_A1_C\right\rangle+\lambda_{3}\left|1_B1_A0_C\right\rangle+\lambda_{4}\left|1_B1_A1_C\right\rangle.\label{nannnannda2}
\end{equation}
We can put \eqref{nannnannda2} in the form of the generalized Schmidt decomposition, after straightforward algebra;
\begin{eqnarray}%\fl
\left|\psi\right\rangle&=&\sqrt{\lambda^2_{3}+\lambda^2_{4}}\left|0'_B0'_A0'_C\right\rangle+\frac{-e^{i\tilde{\varphi}_{5}}(\lambda_{1}\lambda_{3}e^{i\varphi}+\lambda_{2}\lambda_{4})}{\sqrt{\lambda^2_{3}+\lambda^2_{4}}}\left|1'_B0'_A0'_C\right\rangle+\frac{|\lambda_{2}\lambda_{3}-\lambda_{1}\lambda_{4}e^{i\varphi}|}{\sqrt{\lambda^2_{3}+\lambda^2_{4}}}\left|1'_B0_A1_C\right\rangle\nonumber\\
&&+\frac{\lambda_{0}\lambda_{3}}{\sqrt{\lambda^2_{3}+\lambda^2_{4}}}\left|1'_A1'_B0_C\right\rangle+\frac{\lambda_{0}\lambda_{4}}{\sqrt{\lambda^2_{3}+\lambda^2_{4}}}\left|1'_B1'_A1'_C\right\rangle.\label{nannnannda3}
\end{eqnarray}
where $\left|0'_A\right\rangle$, $\left|1'_A\right\rangle$, $\left|0'_B\right\rangle$, $\left|1'_B\right\rangle$, $\left|0'_C\right\rangle$ and $\left|1'_C\right\rangle$ are new basis of the quits $A$, $B$ and $C$.
Note that \eqref{nannnannda3} is the generalized Schmidt decomposition whose order of the qubits is $BAC$.
The coefficients of \eqref{nannnannda3} correspond to the coefficient of \eqref{L2.1}; $\sqrt{\lambda^2_{3}+\lambda^2_{4}}$ corresponds to $\lambda_{0}$, $-e^{i\tilde{\varphi}_{5}}(\lambda_{1}\lambda_{3}e^{i\varphi}+\lambda_{2}\lambda_{4})/\sqrt{\lambda^2_{3}+\lambda^2_{4}}$ corresponds to $\lambda_{1}e^{i\varphi}$, and so on. 
Let us refer to $Q_{\mbox{e}}$ for \eqref{nannnannda3} as $Q^{B}_{\mbox{e}}$.
Because of the definition of $Q_{\mbox{e}}$ and \eqref{nannnannda3},
\begin{eqnarray}
Q^{B}_{\mbox{e}}=\mbox{sgn}\left\{\mbox{Im}\left[\frac{-e^{i\tilde{\varphi}_{5}}(\lambda_{1}\lambda_{3}e^{i\varphi}+\lambda_{2}\lambda_{4})}{|(\lambda_{1}\lambda_{3}e^{i\varphi}+\lambda_{2}\lambda_{4})|}\right]\left(\lambda^2_{3}+\lambda^2_{4}-\frac{K_{5}}{2K_{AC}}\right)\right\}\\
=\mbox{sgn}\left\{\mbox{Im}\left[\frac{-e^{i\tilde{\varphi}_{5}}(\lambda_{1}\lambda_{3}e^{i\varphi}+\lambda_{2}\lambda_{4})}{|(\lambda_{1}\lambda_{3}e^{i\varphi}+\lambda_{2}\lambda_{4})|}\right]\right\}\mbox{sgn}\left(\lambda^2_{3}+\lambda^2_{4}-\frac{K_{5}}{2K_{AC}}\right)\label{nannan0}
\end{eqnarray}
holds.
Then, we can complete the proof by showing that $Q_{\mbox{e}}=Q^{B}_{\mbox{e}}$.
Because \eqref{L2.1} is a positive decomposition and because of \eqref{Qe}, the following two equations hold:
\begin{eqnarray}
&&\mbox{sgn}\left\{\mbox{Im}\left[\frac{-e^{i\tilde{\varphi}_{5}}(\lambda_{1}\lambda_{3}e^{i\varphi}+\lambda_{2}\lambda_{4})}{|(\lambda_{1}\lambda_{3}e^{i\varphi}+\lambda_{2}\lambda_{4})|}\right]\right\}
=\mbox{sgn}\left\{\mbox{Im}[-j_{BC}e^{i\tilde{\varphi}_{5}}(\lambda_{1}\lambda_{3}e^{i\varphi}+\lambda_{2}\lambda_{4})]\right\}\nonumber\\
&&=\mbox{sgn}[\lambda_{1}\lambda_{3}\sin\varphi(-\lambda_{2}\lambda_{3}+\lambda_{1}\lambda_{4}\cos\varphi)-\lambda_{1}\lambda_{4}\sin\varphi(\lambda_{1}\lambda_{3}\cos\varphi+\lambda_{2}\lambda_{4})]\nonumber\\
&&=\mbox{sgn}[-\lambda_{1}\lambda_{2}(\lambda^2_{3}+\lambda^2_{4})\sin\varphi]=-1.\label{nannan2}
\end{eqnarray}
\begin{eqnarray}
\mbox{sgn}\left(\lambda^2_{3}+\lambda^2_{4}-\frac{K_{5}}{2K_{AC}}\right)&=&\mbox{sgn}\left(\frac{K_{AB}}{4\lambda^2_{0}}-\frac{K_{5}}{2K_{AC}}\right)
=\mbox{sgn}\left(\frac{K_{AB}K_{BC}}{2(K_{5}+Q_{\mbox{e}}\sqrt{\Delta_{J}})}-\frac{K_{5}}{2K_{AC}}\right)\nonumber\\
&=&\mbox{sgn}\left[\frac{K_{\mbox{ap}}-K^2_{5}-Q_{\mbox{e}}K_{5}\sqrt{\Delta_{J}}}{2K_{AC}(K_{5}+Q_{\mbox{e}}\sqrt{\Delta_{J}})}\right]
=\mbox{sgn}\left[\frac{-(\sqrt{\Delta_{J}}+Q_{\mbox{e}}K_{5})\sqrt{\Delta_{J}}}{2K_{AC}(K_{5}+Q_{\mbox{e}}\sqrt{\Delta_{J}})}\right]\nonumber\\
&=&\mbox{sqn}\left[\frac{-Q_{\mbox{e}}\sqrt{\Delta_{J}}}{2K_{AC}}\right]=-Q_{\mbox{e}}.\label{nannan1}
\end{eqnarray}
Because of \eqref{nannan0}, \eqref{nannan2} and \eqref{nannan1}, we obtain $Q_{\mbox{e}}=Q^{B}_{\mbox{e}}$.
$\Box$

\section{Derivation of \eqref{revise11} and \eqref{revise4}}
In the present section, we derive \eqref{revise11} and \eqref{revise4}.
Because the decomposition \eqref{cdef} is a decomposition for GHZ-type states, hereafter we assume that  $\tau_{ABC}\ne0$.
Because of $\tau_{ABC}=4\lambda^2_{0}\lambda^2_{4}\ne0$, we can transform \eqref{L2.1} as follows:
\begin{eqnarray}
\left|\psi\right\rangle=\lambda_{0}\left|000\right\rangle+\lambda_{1}e^{i\varphi}\left|100\right\rangle+\lambda_{2}\left|101\right\rangle+\lambda_{3}\left|110\right\rangle+\lambda_{4}\left|111\right\rangle\\
=\frac{1}{\sqrt{\frac{4\lambda^2_{4}}{K_{BC}}}}\left(\left|\tilde{0}_{A}\tilde{0}_{B}\tilde{0}_{C}\right\rangle-\frac{\sqrt{K_{AB}K_{AC}}}{2\lambda^2_{0}\sqrt{K_{BC}}}e^{-i\tilde{\varphi}_{5}}\left|\tilde{1}_{A}\tilde{1}_{B}\tilde{1}_{C}\right\rangle\right),\label{B.2}
\end{eqnarray}
where
\begin{eqnarray}
\left|\tilde{0}_A\right\rangle=\frac{\sqrt{\tau_{ABC}}\left|0_A\right\rangle-C_{BC}e^{-i\tilde{\varphi}_{5}}\left|1_A\right\rangle}{\sqrt{K_{BC}}},\quad\left|\tilde{1}_A\right\rangle=-e^{-i\tilde{\varphi}_{5}}\left|1_A\right\rangle,\\
\left|0'_B\right\rangle=\left|0_B\right\rangle,\quad\left|\tilde{1}_B\right\rangle=\frac{C_{AC}\left|0_B\right\rangle+\sqrt{\tau_{ABC}}\left|1_B\right\rangle}{\sqrt{K_{AC}}},\\
\left|0'_C\right\rangle=\left|0_C\right\rangle,\quad\left|\tilde{1}_C\right\rangle=\frac{C_{AB}\left|0_B\right\rangle+\sqrt{\tau_{ABC}}\left|1_C\right\rangle}{\sqrt{K_{BC}}}.\label{B.5}
\end{eqnarray}
Because of \eqref{B.2}--\eqref{B.5}, the equations \eqref{revise11} and \eqref{revise4} are valid.

%\section*{References}


\begin{thebibliography}{100}
\bibitem{1}D. Deutsch, Proc. R. Soc. Lond. A \textbf{400}, 97 (1985)
\bibitem{2}C. H. Bennett, G. Brassard, C. Crepeau, R. Jozsa, A. Peres, and W. K. Wootters, Phys. Rev. Lett. \textbf{70}, 1895 (1993)
\bibitem{3} C. H. Bennett and S. J. Wiesner, Phys. Rev. Lett. \textbf{69}, 2881 (1992)
\bibitem{4}M. A. Nielsen and I. L. Chuang, Quantum Computation and Quantum Information (Cambridge University Press, Cambridge, 2000)
\bibitem{5}L. P. Hughston, R. Josza, and W. K. Wootters, Phys. Lett. A 1\textbf{83}, 14 (1993)
\bibitem{6}S. Hill and W. K. Wootters, Phys. Rev. Lett. \textbf{78}, 5022 (1997)
\bibitem{7}W. K. Wootters, Phys. Rev. Lett. \textbf{80}, 2245 (1998)
\bibitem{8}G. Vidal and R. F. Werner, Phys. Rev. A \textbf{65}, 032314 (2002)
\bibitem{14}E. Schmidt, Math. Ann. \textbf{63}, 433 (1907)
\bibitem{15}A. Ekert and P. L. Knight, Am. J. Phys. \textbf{63}, 415 (1995)
\bibitem{16}A. Peres, Quantum Theory: Concepts and Methods (Kluwer Academic Publishers, Dordrecht, 1995) 
\bibitem{17}A. Peres, Phys. Lett. A \textbf{202}, 16 (1995)
\bibitem{9}C. H. Bennett, H. J. Bernstein, B. Schumaker, and S. Popescu, Phys. Rev. A \textbf{53}, 2046 (1996)
\bibitem{GHZW}W. D\"{u}r, G.Vidal, and H.I.Cirac, Phys. Rev. A \textbf{62},062314 (2000)
\bibitem{Kraus}B. Kraus, Phys. Rev. Lett. \textbf{104}, 020504 (2010); Phys.
Rev. A \textbf{82}, 032121 (2010).
\bibitem{20}M. A. Nielsen, Phys. Rev. Lett. \textbf{83}, 436 (1999)
\bibitem{21}G. Vidal, Phys. Rev. Lett. \textbf{83}, 1046 (1999)
\bibitem{12}J. Schlienz	and	G. Mahler, Phys. Lett. A \textbf{224}, 39 (1996)
\bibitem{13}N. Linden and S. Popescu, Fortschr. Phys. \textbf{46}, 567 (1998)
\bibitem{10}V. Coffman, J. Kundu, and W. K. Wootters, Phys. Rev. A \textbf{61}, 052306 (2000)
\bibitem{11}A. Miyake, Phys. Rev. A \textbf{67}, 012108 (2003)
\bibitem{18}A. Ac\'{i}n, A. Andrianov, L. Costa, E. Jan\"{e}, J. I. Latorre, and R. Tarrach, Phys. Rev. Lett. \textbf{85} 1560 (2000)  
\bibitem{19}H. A. Carteret, A. Higuchi, and A. Sudbery, J. Math. Phys. \textbf{41}, 7932 (2000)
\bibitem{V}J. I. de Vicente, T. Carle, C. Streitberger, and B. Kraus, Phys. Rev. Lett. 108, 060501 (2012)
\bibitem{60}Each component of these ket vectors represents an eigenstate of the corresponding qubit $A$, $B$ or $C$.  For example, in the case of $\left|101\right\rangle$, which is abbreviation of $\left|1\right\rangle\otimes\left|0\right\rangle\otimes\left|1\right\rangle$, the qubit $A$ is in the eigenstate $\left|1\right\rangle$, the qubit $B$ is in $\left|0\right\rangle$ and the qubit $C$ is in $\left|1\right\rangle$.
We will occasionally use the notation  $\left|1_{A}0_{B}1_{C}\right\rangle$ hereafter.
\bibitem{59}A. Ac\'{i}n, A. Andrianov, E. Jan\"{e} and R. Tarrach, J. Phys. A: Math. Gen. \textbf{34} 6725 (2001)
\bibitem{98}Julio de Vicente, private communication
\bibitem{22}A.  Ac\'{i}n, E. Jan\'{e}, W. D\"{u}r, and G. Vidal, Phys. Rev. Lett. \textbf{85}, 4811 (2000)
\bibitem{32}F. M. Spedalieri, arXiv:quant-ph/0110179v1
\bibitem{24}W. Cui, W. Helwig, and H. K. Lo, Phys. Rev. A \textbf{81}, 012111 (2010)
\bibitem{25}S. Turgut, Y. G\"{u}l, and N. K. Pak, Phys. Rev. A \textbf{81}, 012317 (2010)
\bibitem{26}Zh. L. Cao and M. Yang, J. Phys. B \textbf{36}, 4245 (2003)
\bibitem{27}M. Yang and Zh. L. Cao, Physica A \textbf{337}, 141 (2004)
\bibitem{28}B. Fortescue and H. K. Lo, Phys. Rev. Lett. \textbf{98}, 260501 (2007)
\bibitem{29}B. Fortescue and H. K. Lo, Phys. Rev. A \textbf{78}, 012348 (2008)
\bibitem{30}S. K\i nta\c{s} and S. Turgut, J. Math. Phys. \textbf{51}, 092202 (2010)
\end{thebibliography}
\end{document}